\documentclass[floatfix,showpacs,prb,aps,twocolumn]{revtex4}

\usepackage{bm}
\usepackage{amsmath}
\usepackage{amssymb}
\usepackage{latexsym}
\usepackage{amsfonts}
\usepackage{epsfig}
\usepackage{color}

\newcommand{\ket}[1]{\left|#1\right>}
\newcommand{\bra}[1]{\left<#1\right|}

\newcommand{\expval}[1]{\left< #1 \right>}
\newcommand{\nn}{\nonumber\\}

\newcommand{\f}[1]{\mbox{\boldmath$#1$}}

\newcommand{\bea}{\begin{eqnarray}}
\newcommand{\ea}{\end{eqnarray}}
\newcommand{\eea}{\end{eqnarray}}
\newcommand{\ord}{{\cal O}}
\newcommand{\abs}[1]{{\left| #1 \right|}}
\newcommand{\trace}[1]{{\rm Tr}\left\{ #1 \right\}}

\newcommand{\HS}{H_{\rm S}}
\newcommand{\HI}{H_{\rm SB}}
\newcommand{\HB}{H_{\rm B}}

\definecolor{grey}{rgb}{0.5, 0.5, 0.5}
\definecolor{dgreen}{rgb}{0.0, 0.5, 0.0}
\definecolor{violet}{rgb}{0.5, 0.0, 0.5}
\definecolor{orange}{rgb}{1.0, 0.5, 0.0}

\begin{document}

\title{Transport Statistics of Interacting Double Dot Systems:\\
Coherent and Non-Markovian Effects}

\author{Gernot Schaller}
\email{gernot.schaller@tu-berlin.de}
\author{Gerold Kie{\ss}lich}
\email{gerold.kiesslich@tu-berlin.de}
\author{Tobias Brandes}

\affiliation{Institut f\"ur Theoretische Physik, Hardenbergstra{\ss}e 36,
Technische Universit\"at Berlin, D-10623 Berlin, Germany}

\begin{abstract}
We formalize the derivation of a generalized coarse-graining $n$-resolved master equation by introducing a 
virtual detector counting the number of transferred charges in
single-electron transport. Our approach
enables the convenient inclusion of coherences and Lamb shift in
counting statistics.
As a Markovian example with Lindblad-type density matrices, we consider the Born-Markov-Secular (BMS) approximation 
which is a special case of the non-Markovian dynamical coarse graining (DCG) approach.
For illustration we consider transport 
through two interacting levels that are either serially or parallelly coupled to two leads 
held at different chemical potentials.
It is shown that the coherences can strongly influence the (frequency-dependent) transport
cumulants: In the serial case the neglect of coherences would lead to
unphysical currents through disconnected conductors. Interference
effects in the parallel setup can cause strong current suppression with
giant Fano factors and telegraph-like distribution functions of
transferred electrons, which is not found without coherences. 
We demonstrate that with finite coarse graining times coherences are
automatically included and, consequently, the shortcomings of the BMS
approximation are resolved.
\end{abstract}


\maketitle

%%%%%%%%%%%%%%%%%%%%%%%%%%%%%%%%%%%%%%%%%%%%%%%%%%%%%%%%%%%%%%%%%%%%%%%%%%%%%%%%%%%%
%%%%%%%%%%%%%%%%%%%%%%%%%%%%%%%%%%%%%%%%%%%%%%%%%%%%%%%%%%%%%%%%%%%%%%%%%%%%%%%%%%%%
%%%%%%%%%%%%%%%%%%%%%%%%%%%%%%%%%%%%%%%%%%%%%%%%%%%%%%%%%%%%%%%%%%%%%%%%%%%%%%%%%%%%
%%%%%%%%%%%%%%%%%%%%%%%%%%%%%%%%%%%%%%%%%%%%%%%%%%%%%%%%%%%%%%%%%%%%%%%%%%%%%%%%%%%%
\section{Introduction}
%%%%%%%%%%%%%%%%%%%%%%%%%%%%%%%%%%%%%%%%%%%%%%%%%%%%%%%%%%%%%%%%%%%%%%%%%%%%%%%%%%%%
%%%%%%%%%%%%%%%%%%%%%%%%%%%%%%%%%%%%%%%%%%%%%%%%%%%%%%%%%%%%%%%%%%%%%%%%%%%%%%%%%%%%
%%%%%%%%%%%%%%%%%%%%%%%%%%%%%%%%%%%%%%%%%%%%%%%%%%%%%%%%%%%%%%%%%%%%%%%%%%%%%%%%%%%%
%%%%%%%%%%%%%%%%%%%%%%%%%%%%%%%%%%%%%%%%%%%%%%%%%%%%%%%%%%%%%%%%%%%%%%%%%%%%%%%%%%%%

The number of single electrons transferred
stochastically through a small conductor  in a given time interval obeys a statistics which is
specific to the underlying transport process and to the details of the conductor
\cite{LEV93,LEV96,JON96,BLA00,NAZ03,LEV04}. The deviations,
especially for uncorrelated charge transfer in single tunnel junctions
(Poissonian statistics), from the Gaussian distribution, however, are tiny and
merely become visible in the tails of the distribution.
Nevertheless, the exploration of the so-called full counting
statistics (FCS) has
established an active subfield of mesoscopic transport in recent years.
Although theory work on FCS in mesoscopic transport is still highly dominating the
field, several stimulating experiments were reported recently  
\cite{LU03,REU03,FUJ04,BOM05,BYL05,GUS06,FUJ06a,FRI07,SUK07,FLI09,ZHA09}.
For example, the measurement of up to the 15th-order cumulant of tunneling
through a single quantum dot (QD) and the observation of universal
oscillations in FCS \cite{FLI09} indicate ongoing activities.  

%%%%%%%%%%%%%%%%%%%%%%%%%%%%%%%%%%%%%%%%%%%%%%%%%%%%%%%%%%%%%%%%%%%%%%%%%%%%%%%%%%%%
%%%%%%%%%%%%%%%%%%%%%%%%%%%%%%%%%%%%%%%%%%%%%%%%%%%%%%%%%%%%%%%%%%%%%%%%%%%%%%%%%%%%

\subsection*{Some recent theoretical work on counting statistics}

The theoretical study of FCS in quantum transport is mostly based on the computation of its
cumulant generating function. This turned out to be more convenient
for practical purposes rather than the direct calculation of the distribution function. 
Various methods were developed, e.g., the Levitov-Lesovik formula in the
S-matrix formalism \cite{LEV93,LEV96}, diagonalization of Liouvillians
of Markovian master equations \cite{BAG03a} with the
generalization to frequency-dependent FCS by one of the authors in Ref.~\cite{EMA07}, by means of Stochastic Path
Integral formulation for classical stochastical networks \cite{PIL03,JOR04}, {\em via} a charge representation
method \cite{RAM04}, by an effective field-theory \cite{GUT05}, with nonequilibrium-Greens functions
\cite{GOG06}, in a time-dependent  Levitov-Lesovik approach \cite{SCH07a}, 
by a wave-packet formalism \cite{HAS08}, or
for time-dependent FCS through the formulation by positive-operator-valued measure \cite{BED08}.

The FCS for non-Markovian transport with generalized
master equations were discussed in Ref.~\cite{BRA06} and in
Ref.~\cite{FLI08} wherein the authors introduced an iterative scheme to
compute the cumulants. The interrelation between waiting time distributions and FCS for
single-particle transport were studied in Ref.~\cite{BRA08}.
Esposito {\em et al.} \cite{ESP07} have shown that the FCS for
tunneling through tunnel junctions obeys a fluctuation theorem
relating distribution functions for forward and backward bias voltages.

FCS and noise can be utilized as an important diagnostic probe of
quantum coherence and decoherence mechanisms
\cite{PAL04,FOE05,KIE05a,GRO06,WAN07,WEL08,URB08,URB09}.
In principle, it turns out that the
higher-order cumulants are very sensitive to coherent effects, as
explicitly shown experimentally  and theoretically for the
second-order cumulant (noise) in Ref.~\cite{KIE07b}.\\

The counting statistics of transport through single QDs with attached
single phonon mode (nanoelectromechanical systems, Anderson-Holstein
model) were explored in
Refs.~\cite{FLI05,KOC05a,BEN08,AVR09,SCH09b,HAU09} and shows that the
inelastic scattering processes strongly modify the current fluctuations.

Further system-based theory on FCS can be found for bistable systems
\cite{JOR04a} (generic FCS for telegraph current signals by means of
stochastic-path-integrals), for Andreev scattering in an asymmetric
chaotic cavity \cite{VAN05}, for a QD in the Kondo regime \cite{KOM05},
for B{\"u}ttikers voltage and dephasing probes by Pilgram {\em et al.}
\cite{PIL06}, for spin transfer through ultra-small quantum dots \cite{SCH07}, for transport through a
molecular quantum dot magnet \cite{IMU07}, for chaotic cavities with many
open channels \cite{NOV07}, for Andreev reflection
\cite{DUA09}, and for
spin-transport with ferromagnetic leads \cite{LIN09}.

Belzig studied the FCS of super-Poissonian electron transfer (bunching) in a
single multilevel QD caused by dynamical channel blockade \cite{BEL05}.

The combined counting statistics of transferred electrons and emitted
photons in a single QD reveals crossovers between bunching and
anti-bunching  statistics for both species \cite{SAN07,SAN08a}.

%%%%%%%%%%%%%%%%%%%%%%%%%%%%%%%%%%%%%%%%%%%%%%%%%%%%%%%%%%%%%%%%%%%%%%%%%%%%%%%%%%%%
%%%%%%%%%%%%%%%%%%%%%%%%%%%%%%%%%%%%%%%%%%%%%%%%%%%%%%%%%%%%%%%%%%%%%%%%%%%%%%%%%%%%

\subsection*{Master equations for single-electron transport}

Since we utilize a quantum master equation (ME) we
will provide a brief survey of related previous work.

A common starting point for a systematic perturbative
description of single-electron transport through localized states attached to leads (QDs, molecules, short carbon
nanotubes, {\em et cetera}) is the non-Markovian ME with Born
approximation (system-bath factorization) of the 
reduced density matrix. The assumption that the time evolution of the system
state only depends on the present state  leads to the
Wangsness-Bloch Redfield approach \cite{WAN53,BLO57,RED65,BRE02}. 
A further Markov approximation yields a 
ME which is widely applied in the transport context e.g. in
\cite{LI05,BRA05a,KAI06,HAR06,TIM08,KIE09}.
A well-known shortcoming of this  approximation is
that the positivity of the density matrix is not guaranteed, see
e.g. Ref.~\cite{WHI08} and references therein. 
To circumvent that, a secular approximation can be carried out and the ME
acquires a Lindblad form \cite{LIN76}. 
Another way to avoid the non-positivity is the so-called
singular-coupling limit where the secular approximation is not necessary
\cite{BRE02,SCH09}. Here, we will present another alternative based on
(dynamical) coarse graining.

A different access to the dynamics of the system density matrix is
provided by the Keldysh-contour formulation of K{\"o}nig {\em et al.}
\cite{KOE96,KOE96a}. This method allows for a systematic diagrammatic
expansion in the tunnel coupling. The noise for cotunneling though a
single QD were studied in Ref.~\cite{THI04a}.
The FCS is obtained here by putting the
counting fields ($e^{i\chi}$) in the tunnel Hamiltonian by hand \cite{URB08}.

Microscopic rate equations for transport through coupled QDs were derived in
Refs.~\cite{STO96,GUR96c}. These works are considered to be benchmarks
for the study of serially coupled QDs since the MEs turn out to be
exact for infinite bias. For the first time, Gurvitz {\em et al.}
\cite{GUR96c} have been formulated the ME such that the system density
matrix is resolved with respect to the number of transferred
electrons. The FCS is then readily obtained.

In Ref.~\cite{PED05} Pedersen and Wacker
include broadening effects in the ME description which become
important for finite bias. 
They take into account the
time evolution of coherences between different $k$-states in the leads
due to the tunneling processes. The numerical evaluation of the
$k$-resolved dynamics becomes rather expensive. However, the results agree 
with state-of-the-art methods \cite{PED07}. For that technique FCS
and noise is not considered yet.

Recently, Leijnse and Wegewijs \cite{LEI08} reported on a ME
approach for the reduced density using a Liouville-space
perturbation theory. They systematically expand an effective
Liouvillian in Laplace space with respect to the tunnel coupling.
The FCS in that framework has been studied in Ref.~\cite{EMA09}.

%%%%%%%%%%%%%%%%%%%%%%%%%%%%%%%%%%%%%%%%%%%%%%%%%%%%%%%%%%%%%%%%%%%%%%%%%%%%%%%%%%%%
%%%%%%%%%%%%%%%%%%%%%%%%%%%%%%%%%%%%%%%%%%%%%%%%%%%%%%%%%%%%%%%%%%%%%%%%%%%%%%%%%%%%

\subsection*{This work}

We address the single-electron transport through coupled QDs either in
parallel or in serial configuration for arbitrary Coulomb interaction
strengths between the QDs and finite bias voltages. 
We use a ME based on lowest-order tunnel coupling. The 
approach is thus perturbative and cannot capture Kondo effects.
The dynamical coarse
graining (DCG) method \cite{SCH08} prevents non-positive
density matrices. For infinite coarse-graining times we recover the
known Born-Markov-Secular (BMS) approximation whereas for small 
coarse-graining times the exact (non-Markovian \cite{SCH08,ZED09}) dynamics is obtained.
By introducing a virtual detector for the transferred electrons at one
tunnel junction we are able to calculate the time-resolved FCS and the noise
spectrum at one junction. We will show that for both QD configurations the coherences
in the energy-eigenbasis (off-diagonals of the system density matrix) play an important role
and therefore cannot be neglected.
Particularly, in the parallel setup interferences can lead to strong 
exponential current suppression with negative differential conductance
\cite{BRA04,SCH09,DAR09}. This
accompanies with giant super-Poissonian Fano factors, Dicke-like noise
spectra, and broad distribution functions
of transferred electrons caused by the telegraph-like current signal. 
For the serial setup the BMS approximation (vanishing coherences in
the energy eigenbasis) leads to unphysical behavior when the tunnel
coupling between the dots is small and can be fixed
by assuming a finite coarse-graining time.
In our computations the Lamb shift terms (level renormalization \cite{WUN05}) are
always included and their effect is visible when populations
and coherences in the energy-eigenbasis couple.

In Sec.~\ref{Sdetector} we show how we introduce a virtual detector to the
QD system and how we obtain the FCS and the noise spectrum.
The coarse-graining method and the BMS approximation are presented in Sec.~\ref{Scoarsegraining}.
The models, two QDs in series and in
parallel are introduced in Secs.~\ref{Smodelser} and~\ref{Smodelpar}, respectively.
Sec.~\ref{SSequilib} contains the equilibrium results, Sec.~\ref{SSres_ser} the results for the
serial configuration and Sec.~\ref{SSres_par} the results for the parallel setup.
In the Appendices~\ref{Aliouville_ser} and~\ref{Aliouville_par} we provide the DCG
Liouvillians for parallel and serial setups, respectively. 
The evaluation of the Lamb shift terms appears in Appendix~\ref{Alambshift}.

%%%%%%%%%%%%%%%%%%%%%%%%%%%%%%%%%%%%%%%%%%%%%%%%%%%%%%%%%%%%%%%%%%%%%%%%%%%%%%%%%%%%
%%%%%%%%%%%%%%%%%%%%%%%%%%%%%%%%%%%%%%%%%%%%%%%%%%%%%%%%%%%%%%%%%%%%%%%%%%%%%%%%%%%%
%%%%%%%%%%%%%%%%%%%%%%%%%%%%%%%%%%%%%%%%%%%%%%%%%%%%%%%%%%%%%%%%%%%%%%%%%%%%%%%%%%%%
%%%%%%%%%%%%%%%%%%%%%%%%%%%%%%%%%%%%%%%%%%%%%%%%%%%%%%%%%%%%%%%%%%%%%%%%%%%%%%%%%%%%
\section{Counting Statistics by a virtual detector}\label{Sdetector}
%%%%%%%%%%%%%%%%%%%%%%%%%%%%%%%%%%%%%%%%%%%%%%%%%%%%%%%%%%%%%%%%%%%%%%%%%%%%%%%%%%%%
%%%%%%%%%%%%%%%%%%%%%%%%%%%%%%%%%%%%%%%%%%%%%%%%%%%%%%%%%%%%%%%%%%%%%%%%%%%%%%%%%%%%
%%%%%%%%%%%%%%%%%%%%%%%%%%%%%%%%%%%%%%%%%%%%%%%%%%%%%%%%%%%%%%%%%%%%%%%%%%%%%%%%%%%%
%%%%%%%%%%%%%%%%%%%%%%%%%%%%%%%%%%%%%%%%%%%%%%%%%%%%%%%%%%%%%%%%%%%%%%%%%%%%%%%%%%%%

When one wants to describe counting statistics with a master equation approach, one is first
faced with the problem that the number of tunneled particles is actually a bath and not a system
observable.
In addition, for systems with coherences it appears not so trivial to identify matrix elements of 
the Liouville superoperator with jump superoperators.
Therefore, we perform the counting statistics by adding a virtual detector with 
infinitely many eigenstates in e.g., the right lead.
Formally, this is done by modifying the tunnel Hamiltonian \cite{DOI07}
\bea
d_i \otimes c_{kR}^\dagger \to d_i \otimes b^\dagger \otimes c_{kR}^\dagger\,,\;\;\;
d_i \otimes c_{kL}^\dagger \to d_i \otimes \f{1} \otimes c_{kL}^\dagger\,,
\eea
where $d_i/d_i^\dagger$ and $c_{k\alpha}/c_{k\alpha}^\dagger$ denote the annihilation/creation operators in
the dot system and the lead $\alpha =L,R$, respectively. The detector excitation operator 
\bea
b^\dagger = \sum_{n=-\infty}^{+\infty} \ket{n+1}\bra{n}
\eea
increases the occupation of the detector by one each time an electron is created in the right lead.
When we formally treat the detector operators $b$ and $b^\dagger$ as system operators, we
may perform the conventional Born-Markov-Secular approximation as presented e.g. in Ref.~\cite{BRE02}
and obtain a Lindblad \cite{LIN76} type master equation.
For example, assuming a two-level dot system  characterized by Fock
states $\vert n_1n_2\rangle$ ($n_i\in\{0,1\}$) the Lindblad form would
operate in an infinite-dimensional Hilbert space spanned by
$\ket{00}\otimes \ket{n}$, $\ket{01}\otimes \ket{n}$, $\ket{10}\otimes
\ket{n}$, and $\ket{11}\otimes \ket{n}$.
We can generally decompose the density matrix in this Hilbert-space as
\bea\label{Edmatsys}
\rho_{\rm S}(t) \equiv \sum_{n,m=-\infty}^{+\infty} \rho^{(nm)}(t) \otimes \ket{n}\bra{m}\,,
\eea
where $\rho^{(nm)}(t)$ act on the dot Hilbert space spanned by $\ket{00}$, $\ket{01}$, $\ket{10}$, $\ket{11}$ and can thus
be represented as $4 \times 4$ matrices.
By taking the ''matrix elements'' $\rho^{(n)}(t) \equiv \rho^{(nn)}(t) = \bra{n} {\rho_{\rm S}}(t) \ket{n}$ we see that the
resulting $n$-resolved density matrices are related {\em via} \cite{WAN07}
\bea\label{Enresolved}
\dot \rho^{(n)} \equiv L_0 \rho^{(n)} + L_+ \rho^{(n-1)} + L_- \rho^{(n+1)}\,,
\eea
which implies that we can ignore dynamics of $\rho^{(nm)}(t)$ for $m \neq n$.
Note that the above approach also captures the small bias range leading to bidirectional transport, 
which leads to the occurrence of both $\rho^{(n+1)}$ and $\rho^{(n-1)}$ in contrast to other 
commonly used $n$-resolved master equations \cite{ELA02,AGU04,GUR05,DON08}.
The generalized Liouvillian superoperator $L_0$ contains transitions between the system and the left lead, whereas 
$L_+$ corresponds to jumps from the system towards the right lead, and {\em vice versa} for $L_-$.
Now, according to the measurement postulate \cite{NIE00} counting $m$ particles at time $t$ in the detector would project the density matrix~(\ref{Edmatsys}) to
\bea
\rho_{\rm S}'(t) = \frac{\rho^{(m)}(t)}{\trace{\rho^{(m)}(t)}} \otimes \ket{m}\bra{m}\,,
\eea
such that we may interpret
\bea\label{Eprobdist}
P_n(t) \equiv \trace{\rho^{(n)}(t)}
\eea
as the probability that $n$ particles will be found in the detector when we measure at time $t$.
Note however that the $n$-resolved density matrices $\rho^{(n)}(t)$ derived in this way are still 
positive semidefinite, since Eqn.~(\ref{Enresolved}) corresponds to a Lindblad
form in a higher-dimensional Hilbert space.

The $n$-resolved master Eqn.~(\ref{Enresolved}) can be Fourier-transformed
$\rho(\chi,t)\equiv \sum_n \rho^{(n)}(t) e^{i n \chi}$, such that we
can consider the Markovian cumulant generating function $S_{\rm MK}(\chi, t)$ for the probability distribution~(\ref{Eprobdist})
\bea\label{Ecumgenfunc}
e^{S_{\rm MK}(\chi, t)} \equiv \sum_n P_n(t) e^{i n \chi} = \trace{e^{L(\chi) t} \bar\rho}\,,
\eea
where $L(\chi)\equiv L_0 + e^{+i \chi} L_+ + e^{-i \chi} L_-$ and the initial density matrix is conventionally chosen
as the stationary state (fulfilling $L(0) \bar \rho = 0$).
If there is a distinct eigenvalue of the Fourier-transformed Liouvillian with a largest real part $\lambda_0(\chi)$, this eigenvalue 
will dominate the evolution for large times $t$
\bea
e^{S_{\rm MK}(\chi, t)} &=&\trace{\rho_0^0 e^{\lambda_0(\chi) t} + \rho_0^1 e^{\lambda_1(\chi) t} + \ldots }\nn
&\approx& e^{\lambda_0(\chi) t} \trace{\rho_0^0}=e^{\lambda_0(\chi) t}\,,
\eea
where $\lambda_0(0) = 0$ (stationary state).
In this limit, the derivatives of the cumulant generating function can be approximately determined from
\bea
\left.(-i \partial_\chi)^k S_{\rm MK}(\chi, t)\right|_{\chi=0} \approx \left.(-i \partial_\chi)^k \lambda_0(\chi)\right|_{\chi=0} t\,.
\eea

For the Markovian stationary current we obtain from the cumulant generating function
\bea\label{Ecurrentmk}
I = -i \trace{L'(0) \bar \rho}\,,
\eea
where conventionally $\bar \rho$ denotes the (normalized) stationary state of $L(0)$, 
see also \cite{AGH06a,BRA06b} for similar expressions.\\
The Markovian finite-frequency noise can be obtained from the second cumulant {\em via} the
MacDonald noise formula \cite{MAC48}
\bea\label{Emcdonald}
S_R(\omega) = \int\limits_0^\infty \omega \sin(\omega t) \frac{d}{dt} 
\left(\expval{n^2(t)}-\expval{n(t)}^2\right) dt\,,
\eea
where the regularization 
\bea
\omega \sin(\omega t) \to
\lim_{\epsilon\to 0}
\left[\omega \sin(\omega t) + \epsilon \cos(\omega t)\right] e^{-\epsilon t}
\eea
is implied \cite{FLI05a}.
It is straightforward to show that the second term in the McDonald formula~(\ref{Emcdonald}) is given by $\expval{n(t)}=I t$ with the
Markovian current~(\ref{Ecurrentmk}) and one can analytically obtain the associated integral.
The first term in Eqn.~(\ref{Emcdonald}) can be written as a Laplace transform, such that one finally obtains
for the frequency-dependent noise
\bea\label{Enoisemk}
S_R(\omega) = \Re \trace{\left[2 L'(0) \frac{1}{i \omega \f{1} + L(0)} L'(0) - L''(0)\right]\bar \rho}\,,\nn
\eea
where $L'(0) \equiv \left.\partial_\chi L(\chi)\right|_{\chi=0}$ and $L''(0)\equiv \left.\partial_\chi^2 L(\chi)\right|_{\chi=0}$, 
see also \cite{AGH06a,BRA06b} for similar relations.
Note that $L(0)$ is singular, such that in order to obtain the zero-frequency limit of the above expression, one may either
evaluate the pseudo-inverse of $L(0)$ or directly deduce it from the dominant eigenvalue
$S(0) = - \lambda_0''(0)$.

%%%%%%%%%%%%%%%%%%%%%%%%%%%%%%%%%%%%%%%%%%%%%%%%%%%%%%%%%%%%%%%%%%%%%%%%%%%%%%%%%%%%
%%%%%%%%%%%%%%%%%%%%%%%%%%%%%%%%%%%%%%%%%%%%%%%%%%%%%%%%%%%%%%%%%%%%%%%%%%%%%%%%%%%%
%%%%%%%%%%%%%%%%%%%%%%%%%%%%%%%%%%%%%%%%%%%%%%%%%%%%%%%%%%%%%%%%%%%%%%%%%%%%%%%%%%%%
%%%%%%%%%%%%%%%%%%%%%%%%%%%%%%%%%%%%%%%%%%%%%%%%%%%%%%%%%%%%%%%%%%%%%%%%%%%%%%%%%%%%
\section{Coarse-Graining and the BMS Approximation}\label{Scoarsegraining}
%%%%%%%%%%%%%%%%%%%%%%%%%%%%%%%%%%%%%%%%%%%%%%%%%%%%%%%%%%%%%%%%%%%%%%%%%%%%%%%%%%%%
%%%%%%%%%%%%%%%%%%%%%%%%%%%%%%%%%%%%%%%%%%%%%%%%%%%%%%%%%%%%%%%%%%%%%%%%%%%%%%%%%%%%
%%%%%%%%%%%%%%%%%%%%%%%%%%%%%%%%%%%%%%%%%%%%%%%%%%%%%%%%%%%%%%%%%%%%%%%%%%%%%%%%%%%%
%%%%%%%%%%%%%%%%%%%%%%%%%%%%%%%%%%%%%%%%%%%%%%%%%%%%%%%%%%%%%%%%%%%%%%%%%%%%%%%%%%%%

%%%%%%%%%%%%%%%%%%%%%%%%%%%%%%%%%%%%%%%%%%%%%%%%%%%%%%%%%%%%%%%%%%%%%%%%%%%%%%%%%%%%
%%%%%%%%%%%%%%%%%%%%%%%%%%%%%%%%%%%%%%%%%%%%%%%%%%%%%%%%%%%%%%%%%%%%%%%%%%%%%%%%%%%%

\subsection{Coarse-Graining Master Equation}

It is well known that for nontrivial systems the conventional Born-Markov approximation scheme does
not necessarily lead to Lindblad-type master equations.
In order to obtain these, an additional secular approximation (termed BMS throughout this paper) is required.
Alternatively, the singular coupling limit also yields Lindblad-type master equations \cite{BRE02,SCH09}.

However, it is also known that Lindblad type master equations can also be obtained using 
coarse-grained time-derivatives \cite{LID01,KNE08}.
The Liouvillian will then depend on the coarse-graining timescale.
For example, one may match the perturbative solution to the equation
\bea\label{Edcg_basic}
\frac{d}{dt} \rho(\tau,t) = L(\tau) \rho(\tau,t)
\eea
for the reduced density matrix (using that $L(\tau)$ is small in the interaction picture)
with the perturbative second-order solution of the von-Neumann equation for the full density matrix in the interaction picture
at coarse-graining time $\tau$
\bea\label{Edefcg}
e^{L(\tau) \tau} \rho_{\rm S}^0 = {\rm Tr_B}\left\{U(\tau) \rho_{\rm S}^0 \otimes \rho_{\rm B}^0 U^\dagger(\tau)\right\}\,.
\eea
In the above equation, $U(\tau)$ denotes the time evolution operator in the interaction picture (including time ordering).
For a decomposition of the interaction Hamiltonian
\bea\label{Ehidecomp}
\HI = \sum_{\alpha=1}^M A_\alpha \otimes B_\alpha = \HI^\dagger
\eea
into system ($A_\alpha$) and bath ($B_\alpha$) operators 
Eqn.~(\ref{Edefcg}) defines the corresponding Liouville superoperator as
\bea\label{Edcg_liouville}
L(\tau) [\f{\rho}] &=& \sum_{\alpha\beta=1}^M \frac{1}{\tau} \int\limits_0^\tau dt_1 dt_2 
\Big[\nn
&&- C_{\bar\alpha\bar\beta}(t_1, t_2) \Theta(t_2-t_1) \f{\rho} \f{A_\alpha^\dagger}(t_1) \f{A_\beta^\dagger}(t_2)\nn
&&- C_{\alpha\beta}(t_1, t_2) \Theta(t_1-t_2) \f{A_\alpha}(t_1) \f{A_\beta}(t_2) \f{\rho}\nn
&&+ C_{\bar\alpha \beta}(t_1, t_2) \f{A_\beta}(t_2) \f{\rho} \f{A_\alpha^\dagger}(t_1) \Big]\,,
\eea
which can be shown to be in Lindblad form \cite{SCH09a}.
In the above equation, the time-dependence arises from the interaction picture and the generalized bath
correlation functions have been introduced as
\bea\label{Ebathcfunc}
C_{\alpha\beta}(t_1,t_2) \equiv {\rm Tr_B}\left\{\f{B_\alpha}(t_1) \f{B_\beta}(t_2) \rho_{\rm B}\right\}\,,\nn
C_{\bar\alpha\beta}(t_1,t_2) \equiv {\rm Tr_B}\left\{\f{B_\alpha^\dagger}(t_1) \f{B_\beta}(t_2) \rho_{\rm B}\right\}\,,\nn
C_{\bar\alpha\bar\beta}(t_1,t_2) \equiv {\rm Tr_B}\left\{\f{B_\alpha^\dagger}(t_1) \f{B_\beta^\dagger}(t_2) \rho_{\rm B}\right\}\,.
\eea
With the relation
$\Theta(x) = \frac{1}{2}\left[1+{\rm sign}(x)\right]$
one may separate the Liouville superoperator~(\ref{Edcg_liouville}) into dissipative terms 
of Lindblad form \cite{LIN76} and non-dissipative terms (Lamb-shift) that can be expressed by a commutator
$L_{\rm LS}(\tau) \rho = -i \left[H_{\rm eff}(\tau), \rho\right]$ with an effective Lamb-shift Hamiltonian $H_{\rm eff}(\tau)= H_{\rm eff}^\dagger(\tau)$.
For equilibrium baths ($\left[\HB, \rho_{\rm B}\right]=0$) the bath correlation functions~(\ref{Ebathcfunc}) will 
only depend on the difference of their time arguments and we may introduce the even ($\gamma_{ij}$) and odd ($\sigma_{ij}$) Fourier transforms
\bea\label{Efourier}
C_{ij}(t) &\equiv& \frac{1}{2\pi} \int\limits_{-\infty}^{+\infty} \gamma_{ij}(\omega) e^{+i\omega t} d\omega\,,\nn
C_{ij}(t) {\rm sgn}(t) &\equiv& \frac{1}{2\pi} \int\limits_{-\infty}^{+\infty} \sigma_{ij}(\omega) e^{+i\omega t} d\omega
\eea
of the bath correlation functions.
With these, the two time integrations in Eqn.~(\ref{Edcg_liouville}) may be performed analytically and just
the one-dimensional integral over $\omega$ remains.

%%%%%%%%%%%%%%%%%%%%%%%%%%%%%%%%%%%%%%%%%%%%%%%%%%%%%%%%%%%%%%%%%%%%%%%%%%%%%%%%%%%%
%%%%%%%%%%%%%%%%%%%%%%%%%%%%%%%%%%%%%%%%%%%%%%%%%%%%%%%%%%%%%%%%%%%%%%%%%%%%%%%%%%%%

\subsection{Coarse-Graining Schemes}

In order to go beyond conventional Markovian approaches, the coarse-graining time $\tau$ must scale
with the physical time \cite{SCH08}.
For example, in order to approximate the short-time dynamics of the exact solution well 
(and thereby also the dynamics of the non-Markovian master equation), $\tau$ should scale linearly with the physical time.
In contrast to conventional Markovian master equations, the dynamical coarse-graining method (DCG) \cite{SCH08,SCH09a}
therefore formally solves all coarse-graining master equations~(\ref{Edcg_basic})
and then fixes the coarse-graining time parameter $\tau=t$ in the solution
\bea
\rho(t) \equiv e^{L(t) \cdot t} \rho_0\,.
\eea
For the pure-dephasing spin-boson model, this choice yields the exact solution for the reduced density matrix.

In addition, one can show analytically, that in the limit $\tau\to\infty$ (an equilibrium bath assumed), the Born-Markov-secular approximation
is reproduced (see \cite{SCH08} for the detailed proof):
The result of the temporal integrations in~(\ref{Edcg_liouville}) may be phrased in terms of band filter functions ${\rm sinc}(x)\equiv\sin(x)/x$,
which converge to Dirac $\delta$ distributions for $\tau\to\infty$, see appendix~\ref{Aliouville_ser} for an example.
Thus, in this limit all integrals collapse and~(\ref{Edcg_liouville}) finally reduces to the BMS Liouvillian.
In the short-time limit, we automatically approach the exact solution (and of course also the non-Markovian master equation) 
by construction and for all times positivity is automatically preserved.

However, in the long-time limit it is not {\em a priori} clear why the coarse-graining time $\tau$ should always scale linearly with the
physical time $t$.
In that case, we would always obtain the BMS stationary state in the long-time limit. 
This state has nice classical properties and yields the exact (non-perturbative) solution for the spin-boson pure
dephasing model, but conflicts with some exact quantum solutions -- as already exemplified by the 
single-resonant level model \cite{SCH09a,ZED09}.
Later-on, we will demonstrate further shortcomings of the BMS approximation.
Especially when one goes beyond the weak-coupling limit, a finite coarse-graining (FCG) time may lead to better results, such that 
one might alternatively investigate dependencies of the form 
\bea\label{Ecoarsegrainingmax}
\tau(t) = \frac{t}{1 + \frac{t}{\tau_{\rm max}}}\,.
\eea
Therefore, when one is only interested in stationary (Markovian) results, one may simply evaluate $L(\tau_{\rm max})$.

%%%%%%%%%%%%%%%%%%%%%%%%%%%%%%%%%%%%%%%%%%%%%%%%%%%%%%%%%%%%%%%%%%%%%%%%%%%%%%%%%%%%
%%%%%%%%%%%%%%%%%%%%%%%%%%%%%%%%%%%%%%%%%%%%%%%%%%%%%%%%%%%%%%%%%%%%%%%%%%%%%%%%%%%%

\subsection{Application to Transport}

It is rather straightforward to apply the coarse-graining method to $n$-resolved master equations:
Eqn.~(\ref{Enresolved}) will appear as
\bea
\dot \rho^{(n)}(\tau,t) &\equiv& L_0(\tau) \rho^{(n)}(\tau,t) + L_+(\tau) \rho^{(n-1)}(\tau,t)\nn
&& + L_-(\tau) \rho^{(n+1)}(\tau,t)\,,
\eea
and we can apply the same arguments that lead to Eqn.~(\ref{Ecumgenfunc}) to define a non-Markovian cumulant generating function
\bea\label{Ecumgenfunc_dcg}
e^{S_{\rm DCG}(\chi, \tau, t)} \equiv \trace{e^{L(\chi,\tau) t} \bar\rho}\,,
\eea
where consistency with the BMS approximation is achieved by $L(0)=\lim\limits_{\tau\to\infty}L(0,\tau)$ and we have $L(0) \bar \rho=0$.
When observables (cumulants) should be calculated, the actual scaling of the coarse-graining time with the physical time is of importance, 
since from the above equation we can deduce the time-dependent cumulants by taking derivatives with respect to $\chi$ and $t$.
For example, we will refer to 
\bea
I(t) &\equiv& \frac{d}{dt} \expval{n(\tau(t), t)}\nn
&=& \frac{d}{dt} \left.(-i \partial_\chi) S_{\rm DCG}(\chi, \tau(t), t)\right|_{\chi=0}
\eea
as the time-dependent current.
When $\tau$ is fixed (Markovian case), this current becomes time-independent and when $\tau\to\infty$, we obtain the
BMS current.

%%%%%%%%%%%%%%%%%%%%%%%%%%%%%%%%%%%%%%%%%%%%%%%%%%%%%%%%%%%%%%%%%%%%%%%%%%%%%%%%%%%%
%%%%%%%%%%%%%%%%%%%%%%%%%%%%%%%%%%%%%%%%%%%%%%%%%%%%%%%%%%%%%%%%%%%%%%%%%%%%%%%%%%%%

\subsection{Decomposing the Fock space}

It should be noted that at first sight fermionic hopping terms appear incompatible with the tensor-product representation
of the interaction Hamiltonian~(\ref{Ehidecomp}):
Operators acting on different Hilbert spaces commute by construction, whereas 
in the fermionic hopping terms the dot and lead operators anticommute 
(this feature is also exploited in the exact solutions \cite{SCH09a,ZED09}).
However, we can map such hopping terms to a tensor-product representation.
From the original anticommuting fermionic operators (denoted with overbars), such a 
representation can be obtained from the decomposition
$\bar d_1 = \sigma_1^+ \otimes \f{1} \otimes \f{1}$,
$\bar d_2 = \sigma_1^z \otimes \sigma_2^+ \otimes \f{1}$, and
$\bar c_{ka} = \sigma_1^z \otimes \sigma_2^z \otimes c_{ka}$,
where $\sigma^+=\frac{1}{2}\left(\sigma^x+i \sigma^y\right)$, 
which is similiar to a Jordan-Wigner transform.
In this decomposition, the first two Hilbert spaces refer to the two-dimensional spaces of the first and second system
site, respectively, and the last Hilbert space is simply the (infinite-dimensional) Fock space of the leads, within which
the $c_{ka}$ operators obey the usual fermionic anticommutation relations.
For the double dot systems considered here, the Hilbert space of the system is defined as the Fock space with up to
two particles -- spanned by the basis $\ket{00},\ket{01},\ket{10},\ket{11}$.
In this system Hilbert space we can now define new fermionic operators via
$d_1 = - \sigma_1^+ \otimes \sigma_2^z$ and
$d_2 = - \f{1} \otimes \sigma_2^+$.
The generalization of these mappings to systems with more than two sites is straightforward.
Such mappings appear to be usually performed tacitly in the literature, see e.g. \cite{GUR05,WAN07,CON09}.

%%%%%%%%%%%%%%%%%%%%%%%%%%%%%%%%%%%%%%%%%%%%%%%%%%%%%%%%%%%%%%%%%%%%%%%%%%%%%%%%%%%%
%%%%%%%%%%%%%%%%%%%%%%%%%%%%%%%%%%%%%%%%%%%%%%%%%%%%%%%%%%%%%%%%%%%%%%%%%%%%%%%%%%%%
%%%%%%%%%%%%%%%%%%%%%%%%%%%%%%%%%%%%%%%%%%%%%%%%%%%%%%%%%%%%%%%%%%%%%%%%%%%%%%%%%%%%
%%%%%%%%%%%%%%%%%%%%%%%%%%%%%%%%%%%%%%%%%%%%%%%%%%%%%%%%%%%%%%%%%%%%%%%%%%%%%%%%%%%%
\section{Model Hamiltonians}
\subsection{Two Levels in series}\label{Smodelser}
%%%%%%%%%%%%%%%%%%%%%%%%%%%%%%%%%%%%%%%%%%%%%%%%%%%%%%%%%%%%%%%%%%%%%%%%%%%%%%%%%%%%
%%%%%%%%%%%%%%%%%%%%%%%%%%%%%%%%%%%%%%%%%%%%%%%%%%%%%%%%%%%%%%%%%%%%%%%%%%%%%%%%%%%%
%%%%%%%%%%%%%%%%%%%%%%%%%%%%%%%%%%%%%%%%%%%%%%%%%%%%%%%%%%%%%%%%%%%%%%%%%%%%%%%%%%%%
%%%%%%%%%%%%%%%%%%%%%%%%%%%%%%%%%%%%%%%%%%%%%%%%%%%%%%%%%%%%%%%%%%%%%%%%%%%%%%%%%%%%

\begin{figure}[b]
%these replace certain strings in the eps files by the latex symbols
%use with caution
%\psfrag{PSFep1}{$\epsilon_1$}
%\psfrag{PSFep2}{$\epsilon_2$}
%\psfrag{PSFu}{$U$}
%\psfrag{PSFT}{$T_c$}
%\psfrag{PSFfl}{$f_L\left(\omega, \beta, +\frac{V}{2}\right)$}
%\psfrag{PSFfr}{$f_R\left(\omega, \beta, -\frac{V}{2}\right)$}
%\psfrag{PSFtl}{$t_{kL}$}
%\psfrag{PSFtr}{$t_{kR}$}
\includegraphics[width=0.47\textwidth]{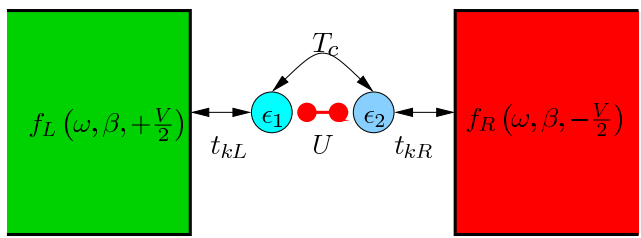}
\caption{\label{Fmodel_ser}[Color Online]
Depending on the bias voltage $V$, particles may travel from left to right or {\em vice versa} through
the effective double dot system.
Setting one tunneling rate ($t_{kL}$, $t_{kR}$, or $T_c$) to zero inhibits the current.
}
\end{figure}

We consider the Hamiltonian\cite{STO96,GUR96c,ELA02,KIE07b}
\bea\label{Emodelser}
H &=& \HS + \HI + \HB\,,\nn
\HS &=& \epsilon_1 d_1^\dagger d_1 + \epsilon_2 d_2^\dagger d_2 + U d_1^\dagger d_1 d_2^\dagger d_2\nn
&& + T_c \left(d_1 d_2^\dagger + d_2 d_1^\dagger\right)\,,\nn
\HB &=& \sum_k \left[ \epsilon_{kL} c_{kL}^\dagger c_{kL} + \epsilon_{kR} c_{kR}^\dagger c_{kR}\right]\,,\nn
\HI &=& \sum_k \left[ t_{kL} d_1 \otimes c_{kL}^\dagger + t_{kR} d_2 \otimes c_{kR}^\dagger + {\rm h.c.}\right]\,,
\eea
where $d_{1/2}^\dagger$ create an electron with different quantum numbers on the dot, respectively,
and $c_{kL/R}^\dagger$ create electrons with momentum $k$ in the left/right lead.
The parameters $\epsilon_i$ denote the single-particle energies, $U$ models the Coulomb interaction, and $T_c$ denotes 
the interdot tunneling rate.
This spinless model could be motivated by a large magnetic field that leads to complete spin polarization
in the leads, such that only one spin would need to be considered, or -- alternatively -- by orbitals where
all tunneling processes are completely symmetric in the electronic spin, such that it 
may be omitted from our considerations.
We assume the symmetric bias case, such that the chemical potential for the left and right leads are kept at
$\mu_L=+V/2$ and $\mu_R=-V/2$, respectively, where $V$ denotes the bias voltage.
The model is depicted in Fig.~\ref{Fmodel_ser}.
The Hilbert space of the system can either be spanned by the localized (particle) basis
$\ket{00},\ket{01},\ket{10},\ket{11}$ or the hybridized eigenbasis of the system Hamiltonian
\bea\label{Eeigenkets}
\ket{E_0} &=& \ket{00}\,,\nn
\ket{E_-} &=& \frac{1}{\sqrt{N}} \left[\left(\delta+\sqrt{\delta^2+4T_c^2}\right) \ket{01}+ 2 T_c \ket{10}\right]\,,\nn
\ket{E_+} &=& \frac{1}{\sqrt{N}} \left[-2 T_c \ket{01}+\left(\delta+\sqrt{\delta^2+4T_c^2}\right) \ket{10}\right]\,,\nn
\ket{E_3} &=& \ket{11}\,,
\eea
with the normalization 
$N=4 T_c^2 + \left(\delta +\sqrt{\delta^2+4T_c^2}\right)^2$ and the single-particle splitting 
\mbox{$\delta \equiv \epsilon_1-\epsilon_2$}.
It can be shown easily that for energetic degeneracy $\delta=0$ the energy eigenbasis remains
non-local even in the limit $T_c\to 0$.
We will see that in the BMS approximation this leads to unphysical artifacts for some parameter values.

With introducing a virtual detector as described in Sec.~\ref{Sdetector}, we identify four coupling operators
in the interaction Hamiltonian
\bea\label{Ecoup_ser}
A_1 &=& d_1 \otimes \f{1}\,,\qquad 		B_1 = \sum_k t_{kL} c_{kL}^\dagger\,,\nn
A_2 &=& d_1^\dagger \otimes \f{1}\,,\qquad	B_2 = \sum_k t_{kL}^* c_{kL}\,,\nn
A_3 &=& d_2 \otimes b^\dagger\,,\qquad 		B_3 = \sum_k t_{kR} c_{kR}^\dagger\,,\nn
A_4 &=& d_2^\dagger \otimes b\,,\qquad		B_4 = \sum_k t_{kR}^* c_{kR}\,.
\eea
The resulting expressions for the bath correlation functions and the Liouville superoperator become
rather lengthy in the general case, but their derivation is outlined in appendix~\ref{Aliouville_ser}.
We assume a continuum of bath modes, such that we may introduce the tunneling rates
\bea\label{Etunnelser}
\Gamma_L(\omega) &\equiv& 2\pi \sum_k \abs{t_{kL}}^2 \delta(\omega -\epsilon_{kL})\,,\nn
\Gamma_R(\omega) &\equiv& 2\pi \sum_k \abs{t_{kR}}^2 \delta(\omega -\epsilon_{kR})\,,
\eea
which we will assume to be approximately constant (flat band limit).
Note however, that for the evaluation of the Lamb-shift terms it is necessary to assume a cutoff, see the
discussion in appendix~\ref{Alambshift}.

Generally, we observe that for finite coarse-graining times $\tau$, the Liouville superoperator~(\ref{Edcg_liouville}) 
couples the six matrix elements
\mbox{$\rho_{00,00},\rho_{-,-},\rho_{+,+},\rho_{11,11},\rho_{-,+},\rho_{+,-}$}
to each other.
In contrast, in the BMS limit ($\tau\to\infty$) and assuming a nondegenerate spectrum, 
the populations \mbox{$\rho_{00,00},\rho_{-,-},\rho_{+,+},\rho_{11,11}$} in the
energy eigenbasis will decouple from the coherences \mbox{$\rho_{-,+},\rho_{+,-}$}, such that it
suffices to consider a $4 \times 4$ Liouville superoperator.
This is a general property of the secular approximation \cite{BRE02} (also rotating
wave approximation \cite{WEL08}).
When the single-charged states become energetically degenerate $\epsilon_1=\epsilon_2$ and $T_c=0$, 
this decoupling does not take place in the BMS limit.

%%%%%%%%%%%%%%%%%%%%%%%%%%%%%%%%%%%%%%%%%%%%%%%%%%%%%%%%%%%%%%%%%%%%%%%%%%%%%%%%%%%%
%%%%%%%%%%%%%%%%%%%%%%%%%%%%%%%%%%%%%%%%%%%%%%%%%%%%%%%%%%%%%%%%%%%%%%%%%%%%%%%%%%%%
%%%%%%%%%%%%%%%%%%%%%%%%%%%%%%%%%%%%%%%%%%%%%%%%%%%%%%%%%%%%%%%%%%%%%%%%%%%%%%%%%%%%
%%%%%%%%%%%%%%%%%%%%%%%%%%%%%%%%%%%%%%%%%%%%%%%%%%%%%%%%%%%%%%%%%%%%%%%%%%%%%%%%%%%%
\subsection{Two Levels in Parallel}\label{Smodelpar}
%%%%%%%%%%%%%%%%%%%%%%%%%%%%%%%%%%%%%%%%%%%%%%%%%%%%%%%%%%%%%%%%%%%%%%%%%%%%%%%%%%%%
%%%%%%%%%%%%%%%%%%%%%%%%%%%%%%%%%%%%%%%%%%%%%%%%%%%%%%%%%%%%%%%%%%%%%%%%%%%%%%%%%%%%
%%%%%%%%%%%%%%%%%%%%%%%%%%%%%%%%%%%%%%%%%%%%%%%%%%%%%%%%%%%%%%%%%%%%%%%%%%%%%%%%%%%%
%%%%%%%%%%%%%%%%%%%%%%%%%%%%%%%%%%%%%%%%%%%%%%%%%%%%%%%%%%%%%%%%%%%%%%%%%%%%%%%%%%%%

\begin{figure}[b]
%these replace certain strings in the eps files by the latex symbols
%use with caution
%\psfrag{PSFep1}{$\epsilon_1$}
%\psfrag{PSFep2}{$\epsilon_2$}
%\psfrag{PSFu}{$U$}
%\psfrag{PSFfl}{$f_L\left(\omega, \beta, +\frac{V}{2}\right)$}
%\psfrag{PSFfr}{$f_R\left(\omega, \beta, -\frac{V}{2}\right)$}
%\psfrag{PSFtl1}{$t_{kL}^1$}
%\psfrag{PSFtl2}{$t_{kL}^2$}
%\psfrag{PSFtr1}{$t_{kR}^1$}
%\psfrag{PSFtr2}{$t_{kR}^2$}
\includegraphics[width=0.47\textwidth]{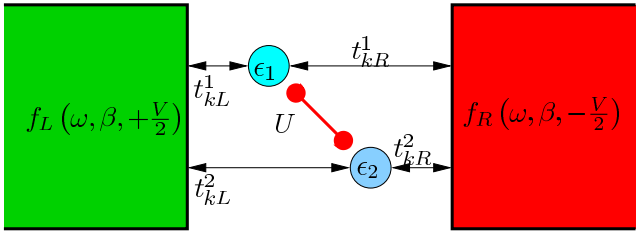}
\caption{\label{Fmodel_par}[Color Online]
Depending on the bias voltage $V$, particles may travel from left to right or {\em vice versa} through
the effective double dot system.
For negligible Coulomb interaction $U$ one has two independent channels.
}
\end{figure}

We consider the Hamiltonian (compare also e.g. \cite{BRA04,BRA06b,WAN07,WEL08,SCH09})
\bea\label{Emodelpar}
H &=& \HS + \HI + \HB\,,\nn
\HS &=& \epsilon_1 d_1^\dagger d_1 + \epsilon_2 d_2^\dagger d_2 + U d_1^\dagger d_1 d_2^\dagger d_2\,,\nn
\HB &=& \sum_k \left[ +\epsilon_{kL} c_{kL}^\dagger c_{kL} + \epsilon_{kR} c_{kR}^\dagger c_{kR}\right]\,,\nn
\HI &=& \sum_k \Big[ t_{kL}^1 d_1^\dagger \otimes c_{kL} + t_{kL}^2 d_2^\dagger \otimes c_{kL}\nn
&&+ t_{kR}^1 d_1^\dagger \otimes c_{kR} + t_{kR}^2 d_2^\dagger \otimes c_{kR} + {\rm h.c.}\Big]\,,
\eea
where we now have four tunneling rates instead of two for the serial model~(\ref{Emodelser}).
As before, we omit the spin from our considerations.
Note that in this model, there is no interdot hopping, 
but the model becomes nontrivial due to the Coulomb interaction term.
The model is depicted in Fig.~\ref{Fmodel_par}.

With introducing a virtual detector as in Sec.~\ref{Sdetector} one can identify eight 
coupling operators in the interaction Hamiltonian
\bea\label{Ecoup_par}
A_1 &=& d_1         \otimes b^\dagger\,,\qquad B_1 = \sum_k t_{kR}^{1*} c_{kR}^\dagger\,,\nn
A_2 &=& d_1^\dagger \otimes b\,,	\qquad B_2 = \sum_k t_{kR}^{1} c_{kR}\,,\nn
A_3 &=& d_2         \otimes b^\dagger\,,\qquad B_3 = \sum_k t_{kR}^{2*} c_{kR}^\dagger\,,\nn
A_4 &=& d_2^\dagger \otimes b\,,	\qquad B_4 = \sum_k t_{kR}^{2} c_{kR}\,,\nn
A_5 &=& d_1         \otimes \f{1}\,,	\qquad B_5 = \sum_k t_{kL}^{1*} c_{kL}^\dagger\,,\nn
A_6 &=& d_1^\dagger \otimes \f{1}\,,	\qquad B_6 = \sum_k t_{kL}^{1} c_{kL}\,,\nn
A_7 &=& d_2         \otimes \f{1}\,,	\qquad B_7 = \sum_k t_{kL}^{2*} c_{kL}^\dagger\,,\nn
A_8 &=& d_2^\dagger \otimes \f{1}\,,	\qquad B_8 = \sum_k t_{kL}^{2} c_{kL}\,.
\eea
With introducing the continuum tunneling rates
\bea\label{Etunnelpar}
\Gamma_{L,1/2}(\omega) &\equiv& 2\pi \sum_k \abs{t_{kL}^{1/2}}^2
\delta(\omega -\epsilon_{kL})\,,\nn
\Gamma_{R,1/2}(\omega) &\equiv& 2\pi \sum_k \abs{t_{kR}^{1/2}}^2
\delta(\omega -\epsilon_{kR})\,,\nn
\gamma_{R/L}(\omega) &\equiv& 2\pi \sum_k t_{kR/L}^{1*} t_{kR/L}^2
\delta(\omega -\epsilon_{kR/L})
\eea
we outline the derivation of the explicit Liouville superoperator in appendix~\ref{Aliouville_par}.
Note that now some tunneling rates may assume complex values.
In the flat band limit, we assume them as frequency independent, which implies $\abs{\gamma_L}^2 = \Gamma_{L1} \Gamma_{L2}$ and
$\abs{\gamma_R}^2 = \Gamma_{R1}\Gamma_{R2}$.

Generally, we find that the Liouville superoperator couples the six matrix elements
\mbox{$\rho_{00,00},\rho_{01,01},\rho_{10,10},\rho_{11,11},\rho_{01,10},\rho_{10,01}$}.
However, in the BMS limit ($\tau\to\infty$) the populations \mbox{$\rho_{00,00},\rho_{01,01},\rho_{10,10},\rho_{11,11}$}
will -- for lifted energetic degeneracy  $\epsilon_1\neq\epsilon_2$ -- decouple from the coherences \mbox{$\rho_{01,10},\rho_{10,01}$}.
For energetic degeneracy however, this decoupling does not take place in the BMS limit.
The nondegenerate BMS case will be termed rate equation ($4 \times 4$ Liouville superoperator), whereas we will refer to the degenerate
BMS case as quantum master equation ($6 \times 6$ Liouville superoperator).

Thus, one can see that the BMS Liouvillian behaves discontinuously with respect to the Hamiltonian parameters, which is of course also 
reflected in observables.
We show in appendix~\ref{Alambshift} how one obtains the Lamb-shift terms (odd Fourier transform) for Lorentzian-shaped
bands.

%%%%%%%%%%%%%%%%%%%%%%%%%%%%%%%%%%%%%%%%%%%%%%%%%%%%%%%%%%%%%%%%%%%%%%%%%%%%%%%%%%%%
%%%%%%%%%%%%%%%%%%%%%%%%%%%%%%%%%%%%%%%%%%%%%%%%%%%%%%%%%%%%%%%%%%%%%%%%%%%%%%%%%%%%
%%%%%%%%%%%%%%%%%%%%%%%%%%%%%%%%%%%%%%%%%%%%%%%%%%%%%%%%%%%%%%%%%%%%%%%%%%%%%%%%%%%%
%%%%%%%%%%%%%%%%%%%%%%%%%%%%%%%%%%%%%%%%%%%%%%%%%%%%%%%%%%%%%%%%%%%%%%%%%%%%%%%%%%%%
%%%%%%%%%%%%%%%%%%%%%%%%%%%%%%%%%%%%%%%%%%%%%%%%%%%%%%%%%%%%%%%%%%%%%%%%%%%%%%%%%%%%
%%%%%%%%%%%%%%%%%%%%%%%%%%%%%%%%%%%%%%%%%%%%%%%%%%%%%%%%%%%%%%%%%%%%%%%%%%%%%%%%%%%%
\section{BMS Equilibrium Results and Fluctuation-Dissipation relations}\label{SSequilib}

For a system in equilibrium (either obtained by choosing identical chemical potentials in 
both reservoirs $\mu_L=\mu_R=\mu$ or by turning off one coupling (e.g. $\Gamma_L=0$ and $\mu_R=\mu$ or 
$\Gamma_R=0$ and $\mu_L=\mu$) in the serial case) we obtain equilibration 
of both temperature and chemical potentials with that of the reservoir.
This holds for both models~(\ref{Emodelser}) and~(\ref{Emodelpar}).
More explicitly, when the reservoir is in the state
\bea
\rho_{\rm B}=\frac{e^{-\beta\left(\HB - \mu N_{\rm B}\right)}}{Z_{\rm B}}\,,
\eea
the resulting BMS Liouville superoperators has the steady state
\bea
\bar\rho_{\rm S} = \frac{e^{-\beta\left(\HS - \mu N_{\rm S}\right)}}{Z_{\rm S}}\,,
\eea
where $N_{\rm S}=d_1^\dagger d_1 + d_2^\dagger d_2$ is the system particle number operator.
The equilibration of both temperature $\beta$ and chemical potential
$\mu$ between system and reservoir for these systems is a remarkable result, since to
our knowledge only equilibration  of temperatures has been shown so
far \cite{BRE02}.

In all analytically accessible cases, we can verify the
Johnson-Nyquist (fluctuation-fissipation) relation \cite{JOH28,NYQ28}
\bea
S(0)\Big|_{V=0} = \frac{2}{\beta} \frac{d I}{dV}\Big|_{V=0}\, ,
\eea
which requires bi-directional $n$-resolved master equations of the form~(\ref{Enresolved})
and appears to be a general feature of the BMS approximation \cite{CON09}.
For zero bias voltage $V=0$ we have shown as a sanity check that the resulting current vanishes.

A fluctuation-dissipation relation for nonlinear transport based on detailed balance has recently been derived in Ref.~\cite{FOE08}:
$\big\langle e^{\vec{A}\cdot\vec{Q}}\big\rangle =1$
with $\quad\vec{A}\equiv e\beta\{ V_L,V_R\}=e\beta\{ V/2,-V/2\}$ and $\vec{Q}\equiv \{n_L,n_R\}$ for  two terminals and symmetric bias.
We have checked this relation for a single noninteracting level and find it satisfied within the BMS approximation 
only up to third order in $V$. 
However, since we only consider counting at the right junction the check for the interacting systems is beyond the scope of the present work.

%%%%%%%%%%%%%%%%%%%%%%%%%%%%%%%%%%%%%%%%%%%%%%%%%%%%%%%%%%%%%%%%%%%%%%%%%%%%%%%%%%%%
%%%%%%%%%%%%%%%%%%%%%%%%%%%%%%%%%%%%%%%%%%%%%%%%%%%%%%%%%%%%%%%%%%%%%%%%%%%%%%%%%%%%
\section{Transport results: Serial Configuration}\label{SSres_ser}

%%%%%%%%%%%%%%%%%%%%%%%%%%%%%%%%%%%%%%%%%%%%%%%%%%%%%%%%%%%%%%%%%%%%%%%%%%%%%%%%%%%%
\subsection{BMS Rate Equation}

\begin{figure}[t]
%\psfrag{PSFIinf1}{$\frac{\Gamma}{2}$}
%\psfrag{PSFIinf2}{$\color{dgreen}\frac{2 \Gamma T_c^2}{\delta^2+4 T_c^2}$}
%\psfrag{PSFIcb1}{$\frac{\Gamma}{3}$}
%\psfrag{PSFIcb2}{$\color{dgreen}\frac{\Gamma T_c^2}{\delta^2 + 3 T_c^2}$}
%\psfrag{PSFT1}{$\color{blue}\ket{00}\leftrightarrow\ket{E_-},\ket{E_+}$}
%\psfrag{PSFT2}{$\color{blue}\ket{E_+}\leftrightarrow\ket{11}$}
%\psfrag{PSFT3}{$\color{blue}\ket{E_-}\leftrightarrow\ket{11}$}
\includegraphics[width=0.47\textwidth,clip=true]{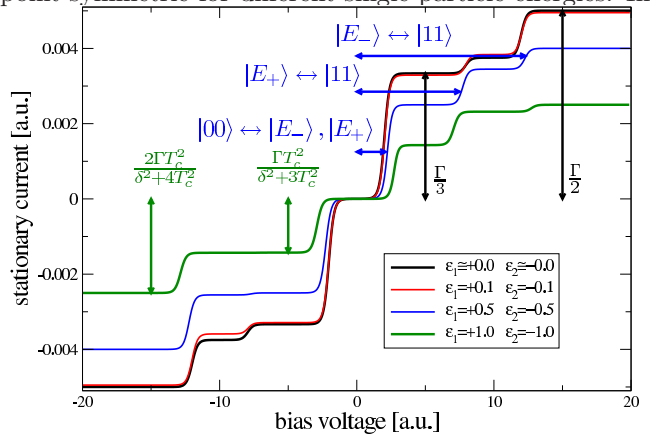}
\caption{\label{Fsercurrent_bias}[Color Online]
Stationary Markovian current {\em versus} bias voltage for the serial double dot for different single-particle energies.
Other parameters have been chosen as $T_c=1.0$, $U=5.0$, $\Gamma_L=\Gamma_R=0.1^2=\Gamma$, $\beta=10.0$.
For degenerate single-particle energies (black solid line) the current is point-symmetric around the origin.
This does not persist for non-degenerate single particle energies and becomes strongly pronounced for large splittings
$\delta\equiv\epsilon_1-\epsilon_2$.
For $\epsilon_1=-\epsilon_2$ one observes three non-vanishing plateau currents for both positive and negative bias voltages 
(given a sufficiently low temperature).
The first current step enables transport by transitions from $\ket{00}\to\ket{E_-}$ and $\ket{00}\to\ket{E_+}$, and at the
second and third current steps the additional channels $\ket{E_+}\to\ket{11}$ and $\ket{E_-}\to\ket{11}$ are opened.
It is visible that with large level splitting (solid green), the second plateau equilibrates with the third one for 
positive bias, whereas it approaches the first plateau for negative bias.
This is a localization effect (for explanations see the text).}
\end{figure}

The BMS current in Coulomb-blockade but high-bias limit (CBHB, formally obtained by the limits
$\lim\limits_{V\to\infty}\lim\limits_{U\to\infty}$ such that always $V \ll U$) equates to
\bea\label{Esercur_cb}
I^{\rm CB} = \frac{\Gamma_L \Gamma_R T_c^2}{\left(\epsilon_1-\epsilon_2\right)^2 \Gamma_L + \left(2 \Gamma_L + \Gamma_R\right) T_c^2}\,,
\eea
which is also found in the numerical solution for the stationary current, see Fig.~\ref{Fsercurrent_bias}.
The above result differs from known results in the literature \cite{GUR96c,STO96,ELA02} by a missing term $\Gamma_L \Gamma_R^2/4$ in the denominator.
This difference is of higher order than the validity of the perturbation theory.
However, this leads to the unphysical artifact that in our case for $\epsilon_1=\epsilon_2$ the current becomes
independent on the interdot hopping rate $T_c$, which contrasts with
the expectation that it should always vanish for 
$T_c\to 0$.
The main reason for this failure of the BMS approximation lies in the use of the non-localized energy eigenbasis 
during the secular approximation procedure: 
Tunneling may occur into non-adjacent eigenstates even when $T_c$ is small, which finally leads to the observed current.
When the secular approximation is not performed (technically, by using a finite maximum coarse-graining time), we will see
that the current vanishes for $T_c\to 0$ as expected, see Sec.~\ref{SSScoarse_graining_results}.
In the actual infinite bias case (formally obtained by the limit
$\lim\limits_{V\to\infty}$ such that we keep finite $U$) we obtain the stationary current
\bea\label{Esercur_ib}
I^\infty = \frac{\Gamma_L \Gamma_R \left(\Gamma_L + \Gamma_R\right) T_c^2}{\left(\epsilon_1-\epsilon_2\right)^2 \Gamma_L \Gamma_R + \left(\Gamma_L + \Gamma_R\right)^2 T_c^2}\,,
\eea
which also displays the same artifact as the Coulomb-blockade current~(\ref{Esercur_cb})) for 
degenerate single-particle energies, but is -- as one would expect -- symmetric under exchange of $L$ and $R$.
This current is also found in the numerical solution, see Fig.
\ref{Fsercurrent_bias}, and differs from \cite{ELA02} by a missing term
$\Gamma_L\Gamma_R(\Gamma_L+\Gamma_R)^2/4$ in the denominator.
The current-voltage characteristics does now exhibit several current steps, since we explicitly allow for a doubly occupied system.
Since we have chosen symmetrized single-particle energies ($\epsilon_2=-\epsilon_1$), the lowest two transport-channels are
not distinguished.
Naturally, the current-voltage characteristics is no longer point-symmetric for different single-particle energies.
In addition, when the splitting is significantly larger than the interdot tunneling rate $T_c$, we see in Fig.~\ref{Fsercurrent_bias} 
(green curve) that one transport channel is strongly suppressed.
This effect can be interpreted {\em via} the localization of the wave functions: 
For $\abs{\epsilon_1-\epsilon_2}\gg T_c$ (large splitting),
the eigenvectors~(\ref{Eeigenkets}) become localized, e.g., for $\epsilon_1>\epsilon_2$ one obtains that 
$\ket{E_+}\to\ket{10}$ and $\ket{E_-}\to\ket{01}$.
When they are above the Fermi surface of their adjacent lead, they may not contribute to transport, such that
only one state ($\ket{E_+}$ for positive bias and $\ket{E_-}$ for negative bias) mediates the transport.
Naturally, this also leads to asymmetries in the voltage-dependent zero-frequency Fano factor, see Fig.~\ref{Fserfano_bias} below.
Note that for extremely large splittings between the single particle energies $\epsilon_i$ 
(constant interdot tunneling rate $T_c$ assumed) the Coulomb-blockade assumption may even
become invalid (not shown), as the energy for the doubly-charged system may fall between the two single-charged energies (not shown).

\begin{figure}[t]
\includegraphics[width=0.47\textwidth,clip=true]{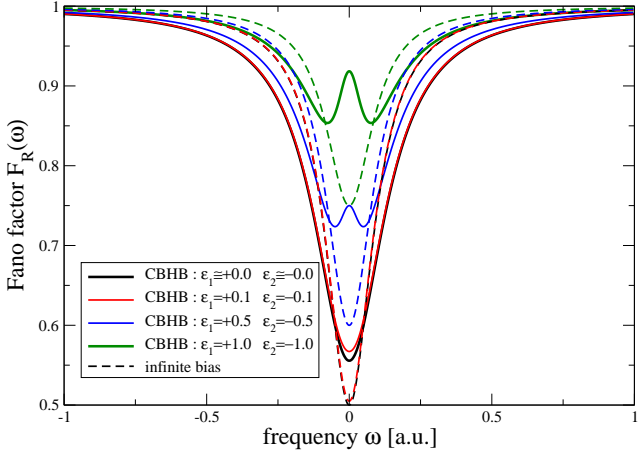}
\caption{\label{Fserfano_omega}[Color Online]
Frequency dependent Fano factor $F_R(\omega)$ for different splittings in Coulomb-blockade high-bias (CBHB, solid) and infinite bias (dashed)
limit {\em versus} $\omega$.
Other parameters have been chosen as $T_c=1.0$, $U=5.0$, $\Gamma_L=\Gamma_R=0.1=\Gamma$, and $\beta=10.0$.
The frequency-dependent Fano factor exhibits two minima that are for large splittings
$\delta\gg T_c$ situated at $\omega^* = \pm \Gamma$ (solid lines).
In the infinite bias limit, this feature is generally lost (dashed lines).
}
\end{figure}

\begin{figure}[t]
\includegraphics[width=0.47\textwidth,clip=true]{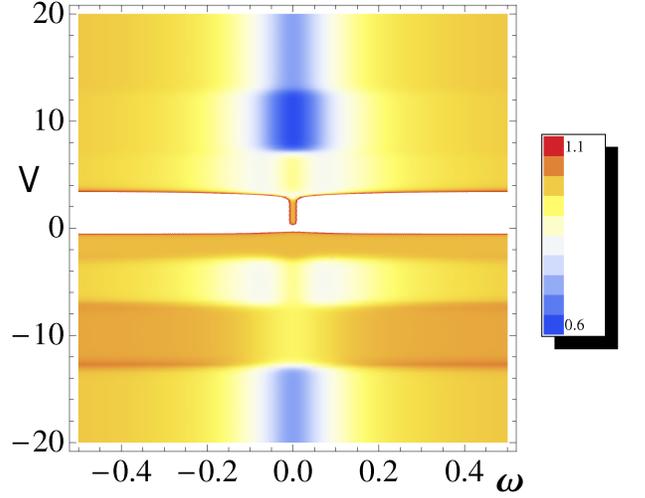}
\caption{\label{Fserfano_omega_bias}[Color Online]
Frequency dependent Fano factor $F_R(\omega)$ for large splitting {\em versus} bias voltage and $\omega$.
Other parameters have been chosen as in Fig.~\ref{Fserfano_omega}, but the
single-particle energies have been fixed at $\epsilon_1=+1.0$ and $\epsilon_2=-1.0$.
In the CBHB regime (e.g. $V=\pm 5$), there are two valleys corresponding to two minima of the Fano factor, 
whereas this feature is lost in the infinite bias regime (e.g. $V=\pm 20$), 
compare also Fig.~\ref{Fserfano_omega}.
}
\end{figure}

From Eqn.~(\ref{Enoisemk}) we may calculate the finite-frequency noise for the right lead
and we obtain for the frequency-dependent Fano factor $F_R(\omega)\equiv S_R(\omega)/\abs{I}$ 
in the CBHB regime the expression
\begin{widetext}
\bea\label{Eserfano_cb}
F^{\rm CB}_{R}(\omega) &=& 1 +
\frac{2 \Gamma_L \Gamma_R T_c^2 \left[\Gamma_R \left(\delta^2 (\Gamma_L - \Gamma_R) - 2 \Gamma_R T_c^2\right) - 2 \left(\delta^2+4 T_c^2\right)\omega^2\right]}
{\left[\delta^2 \Gamma_L \Gamma_R + \Gamma_R \Gamma_{\rm eff} T_c^2\right]^2
+\left(\delta^2 + 4 T_c^2\right) \left[\delta^2 \left(\Gamma_L^2 + \Gamma_R^2\right) + 
\left(\Gamma_{\rm eff}^2 + \Gamma_R^2\right)
T_c^2\right] \omega^2 + \left(\delta^2 + 4 T_c^2\right)^2 \omega^4}\,,
\eea
\end{widetext}
where we have used the short-hand notation $\delta\equiv \epsilon_1-\epsilon_2$ and $\Gamma_{\rm eff}\equiv 2 \Gamma_L + \Gamma_R$.
The zero-frequency limit is depicted in Fig.~\ref{Fserfano_bias}, where the numerical solution coincides with
the above result in the CBHB regime. The
difference between the above zero-frequency Fano factor and the Fano
factor in Ref.~\cite{ELA02} is of the order $\ord{(\Gamma_L\Gamma_R/T_c^2)}$.
It can also be deduced that the zero-frequency limit of Eqn.~(\ref{Eserfano_cb}) 
becomes super-Poissonian (indicating bunching) \cite{BUL00,KIE03b,COT04,THI04,DJU05} 
when the right-associated tunneling rate becomes significantly smaller than the left-associated one
$\Gamma_R < \frac{\Gamma_L}{1+2T_c^2/\delta^2}$, which corresponds to a highly asymmetric situation.
Assuming similar single-particle energies ($\delta=0$) the zero-frequency version of Eqn.~(\ref{Eserfano_cb})
reproduces previous results in the literature \cite{AGH06a}.
Assuming equal tunneling rates $\Gamma_L=\Gamma_R$ we see that the frequency-dependent Fano factor in 
BMS approximation has two minima only if
the splitting between the single-particle energies is larger than the tunneling rate 
($\abs{\delta}=\abs{\epsilon_1-\epsilon_2} \ge \sqrt{\sqrt{5}-2} \abs{T_c} \approx 0.49 \abs{T_c}$)
and is featureless (only one minimum) otherwise.
This is well confirmed in the numerical solution, see the solid lines in Fig.~\ref{Fserfano_omega}.
Even when one has two minima in the finite-frequency noise, Fig.~\ref{Fserfano_omega_bias} demonstrates that
this only happens in the CBHB limit.
For the frequency-dependent Fano factor in the infinite bias limit we obtain
\begin{widetext}
\bea\label{Eserfano_ib}
F^\infty_R(\omega) = 1 
-\frac{2 \Gamma_L \Gamma_R T_c^2 \left[\delta^2 \left(\Gamma_L^2 -
\Gamma_L \Gamma_R + \Gamma_R^2 + \omega^2\right) 
+ T_c^2 \left(\Gamma^2 + 4 \omega^2\right)\right]}
{\left(\delta^2 \Gamma_L \Gamma_R + \Gamma^2 T_c^2\right)^2 + \left(\delta^2 + 4 T_c^2\right) \left[\delta^2 \left(\Gamma_L^2 + \Gamma_R^2\right)+2 \Gamma^2 T_c^2\right] \omega^2 + \left(\delta^2 + 4 T_c^2\right)^2 \omega^4}\,,
\eea
\end{widetext}
where we have again used the abbreviation $\delta\equiv
\epsilon_1-\epsilon_2$ but this time use $\Gamma = \Gamma_L + \Gamma_R$.
The zero-frequency limit is well reproduced in figure
\ref{Fserfano_bias} in the infinite-bias regime. As before, the
difference between the above zero-frequency Fano factor and the Fano
factor in Ref.~\cite{ELA02} is of the order $\ord{(\Gamma_L\Gamma_R/T_c^2)}$.
It is straightforward to show that the zero-frequency version of the Fano factor~(\ref{Eserfano_ib}) may never become super-POoissonian.
In addition, for equal tunneling rates $\Gamma_L=\Gamma_R$ the infinite bias frequency-dependent Fano factor has only one minimum -- regardless of the
splitting of the single particle energies, see the dashed lines in Fig.~\ref{Fserfano_omega}.
\begin{figure}[t]
%\psfrag{PSFfcb}{$\color{blue}1-\frac{4 T_c^4}{\left(\delta^2+3 T_c^2\right)^2}$}
%\psfrag{PSFfinf}{$\color{dgreen}\frac{\delta^2+2T_c^2}{\delta^2+4T_c^2}$}
\includegraphics[width=0.47\textwidth,clip=true]{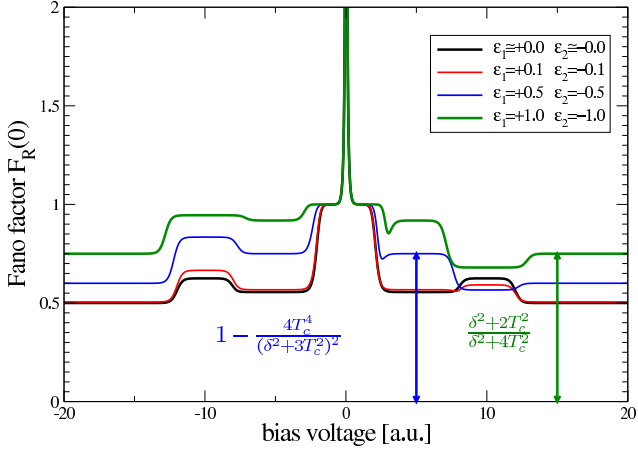}
\caption{\label{Fserfano_bias}[Color Online]
Zero frequency Fano factor {\em versus} bias voltage for different single-particle energies.
Parameters and color coding coincide with Fig.~\ref{Fsercurrent_bias}.
For degenerate single-particle energies (black) we have a symmetric Fano factor.
For large splittings $\delta\equiv\epsilon_1-\epsilon_2$ (green) the equilibration behavior of the Fano plateaus coincides with that of the
current plateaus.
The peak at the origin results from (thermal) Nyquist noise, cf. \cite{AGH06a}.
}
\end{figure}

Finally, Fig.~\ref{Fserfano_omega_bias} demonstrates the smooth transition of the frequency-dependent Fano factor from the
Coulomb-blockade regime towards the infinite-bias limit:
For small positive bias, the frequency-dependent Fano factor becomes slightly super-Poissonian for finite $\omega$ and poissonian
for $\omega=0$, compare also Fig.~\ref{Fserfano_bias}.
In the CBHB regime (solid green line), we observe a sub-Poissonian Fano factor with two minima and in the 
infinite bias regime (dashed green line), the two minima merge into one.

%%%%%%%%%%%%%%%%%%%%%%%%%%%%%%%%%%%%%%%%%%%%%%%%%%%%%%%%%%%%%%%%%%%%%%%%%%%%%%%%%%%%

\subsection{Coarse-Graining Results}\label{SSScoarse_graining_results}

\begin{figure}[t]
\includegraphics[width=0.47\textwidth,clip=true]{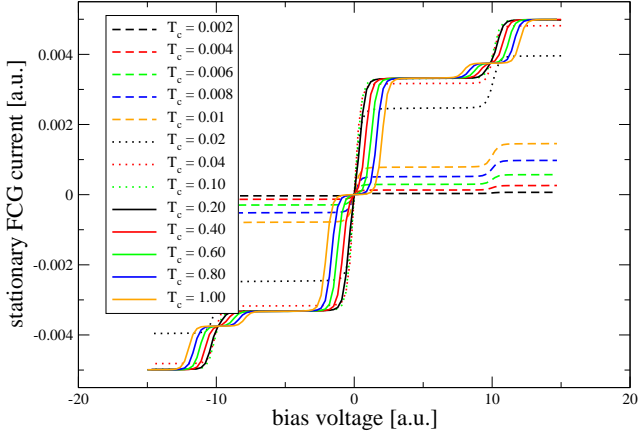}
\caption{\label{FsercurrentFCG_bias}[Color Online]
Plot of the stationary current {\em versus} bias voltage for different interdot tunneling rates $T_c$.
The coarse-graining time has been chosen as $\tau=100$, the inverse of the coupling strengths $\Gamma_L$ and $\Gamma_R$.
For small $T_c$, we observe a quadratic growth of the current magnitude (dashed curves), which quickly saturates.
For larger $T_c$ (solid lines), the current magnitude becomes independent of $T_c$ as also seen in the
BMS approximation for $\epsilon_1=\epsilon_2$, but the level splitting between the single-charged states $\ket{-}$ and 
$\ket{+}$ is slowly resolved (emerging additional current step).
Other parameters were chosen as $\epsilon_1=\epsilon_2=0$, $U=5.0$, $\Gamma_L=\Gamma_R=0.1^2$, and $\beta=10.0$.
}
\end{figure}

For finite coarse-graining times, the Coarse-Graining Liouville superoperator does not resemble the
BMS approximation but nevertheless preserves positivity (Lindblad form), compare the discussion
in Sec.~\ref{Scoarsegraining}.
We have seen that for degenerate single-particle energies $\epsilon_1=\epsilon_2$, the
BMS currents ($\tau\to\infty$) become independent of the interdot tunneling rate $T_c$, compare
equations~(\ref{Esercur_cb}) and~(\ref{Esercur_ib}).
That is, when $T_c\to 0$, one still has a nonvanishing current, which is completely unreasonable, 
as in this limit one has two single levels coupled to different leads.
Note however, that the exact limit $T_c\to 0$ would in this case lead to a different BMS quantum master equation with
a $6 \times 6$ Liouvillian that does not admit any current.
This unphysical behavior is not found for a finite coarse-graining time, see figure
\ref{FsercurrentFCG_bias}.
For a finite coarse-graining time, the stationary current vanishes quadratically as the interdot tunneling rate 
$T_c$ goes to zero, which is consistent with earlier results for the CBHB regime in the literature \cite{GUR96c,STO96}.
For larger interdot tunneling rates, the magnitude of the stationary current quickly saturates, such that the 
only effect of a varying $T_c$ is the increasing level splitting for the single-charged states $\ket{-}$ and $\ket{+}$
(emerging additional current steps).
Therefore, for large $T_c$ we obtain the previous results of the BMS approximation (compare e.g. orange curve in 
Fig.~\ref{FsercurrentFCG_bias} with the black curve in Fig.~\ref{Fsercurrent_bias}).
This behavior is quite general: When the coarse-graining time is larger than the inverse level splitting 
(in our case between the single-charged states), one can show that the coarse-graining results and BMS results coincide.

Regarding the dynamical coarse-graining approach \cite{SCH08,SCH09a}, these findings demonstrate that for the
serial double dot considered here, the coarse-graining time must not be sent to infinity.
To obtain non-Markovian effects, it must for small times scale linearly with the physical time $\tau=t$, but for large times
it must saturate $\tau=\tau_{\rm max}$ in order to avoid artifacts such as currents through disconnected structures.

%%%%%%%%%%%%%%%%%%%%%%%%%%%%%%%%%%%%%%%%%%%%%%%%%%%%%%%%%%%%%%%%%%%%%%%%%%%%%%%%%%%%
%%%%%%%%%%%%%%%%%%%%%%%%%%%%%%%%%%%%%%%%%%%%%%%%%%%%%%%%%%%%%%%%%%%%%%%%%%%%%%%%%%%%
\section{Transport results: Parallel Configuration}\label{SSres_par}

%%%%%%%%%%%%%%%%%%%%%%%%%%%%%%%%%%%%%%%%%%%%%%%%%%%%%%%%%%%%%%%%%%%%%%%%%%%%%%%%%%%%

\subsection{BMS Rate Equation}

\begin{figure}[t]
%\psfrag{PSFIinf}{$\Gamma$}
%\psfrag{PSFIcb}{$\frac{2}{3}\Gamma$}
%\psfrag{PSFT1}{$\color{blue}\ket{00}\leftrightarrow\ket{01},\ket{10}$}
%\psfrag{PSFT2}{$\color{blue}\ket{10}\leftrightarrow\ket{11}$}
%\psfrag{PSFT3}{$\color{blue}\ket{01}\leftrightarrow\ket{11}$}
\includegraphics[width=0.47\textwidth,clip=true]{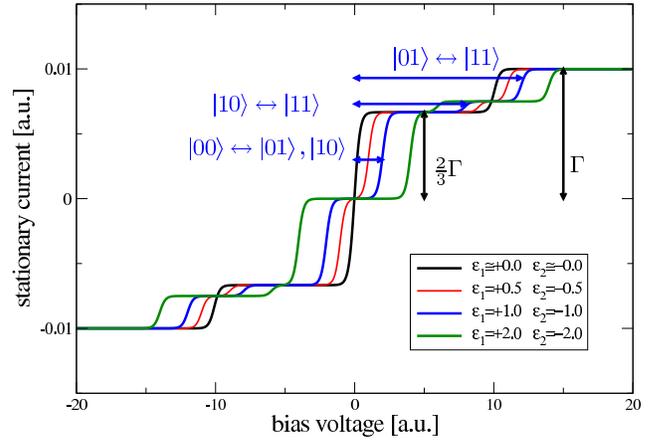}
\caption{\label{Fparcurrent_bias}[Color Online]
Stationary Markovian rate equation current {\em versus} bias voltage for the parallel double dot for different single-particle energies.
Other parameters have been chosen as $U=5.0$, $\Gamma_{L1}=\Gamma_{L2}=\Gamma_{R1}=\Gamma_{R2}=0.1^2=\Gamma$, $\beta=10.0$.
For sufficiently low temperature, the current steps may be found at the (doubled) excitation spectrum of the system Hamiltonian (sequential tunneling limit).
As all tunneling rates are assumed equal, the curves are always point-symmetric around the origin.
For sufficiently large splittings (green curve) and low temperatures one may resolve between different transport channels.
The height of the plateaus is independent of the system energies.
For symmetric single-particle energies one has three plateaus with a non-vanishing current.
With increasing splitting (green), the second plateau broadens thereby annihilating the first one.
This happens when the state $\ket{11}$ becomes energetically more favorable than $\ket{10}$, such that 
the coulomb-blockade assumption is not applicable.
}
\end{figure}

When the single-particle energies are non-degenerate $\epsilon_1 \neq \epsilon_2$, the equations for the populations
decouple from the equations for the coherences, see appendix~\ref{Aliouville_par} for details.
For these rate equations we obtain in the CBHB limit the current
\bea\label{Eparcur_rate_cb}
I^{\rm CB} = \frac{\left(\Gamma_{L1} + \Gamma_{L2}\right) \Gamma_{R1} \Gamma_{R2}}{\Gamma_{L1} \Gamma_{R2} + \Gamma_{L2} \Gamma_{R1} + \Gamma_{R1} \Gamma_{R2}}\,.
\eea
For equal left- and right-associated tunneling rates
($\Gamma_{L1}=\Gamma_{L2}=\Gamma_{L}$,
$\Gamma_{R1}=\Gamma_{R2}=\Gamma_{R}$) we recover the known result \cite{GLA88c}
$2\Gamma_{L}\Gamma_{R}/(2\Gamma_{L}+\Gamma_{R})$.
Clearly, the above current vanishes if just one of the right-associated tunneling rates goes to zero.
This is a Coulomb-blockade effect: Due to the high-bias limit, an electron gets stuck in the orbital with the vanishing
right-associated tunneling rate, compare also Fig.~\ref{Fmodel_par}.
This electron blocks the transport through the other channel by the Coulomb interaction.
The current~(\ref{Eparcur_rate_cb}) is also well confirmed by the numerical solution, see Fig.~\ref{Fparcurrent_bias}.
In contrast, in the actual infinite bias case we obtain the current
\bea\label{Ecurrent_par_infbias}
I^\infty = \frac{\Gamma_{L1} \Gamma_{L2} \left(\Gamma_{R1} + \Gamma_{R2}\right) + \left(\Gamma_{L1} + \Gamma_{L2}\right)\Gamma_{R1} \Gamma_{R2}}
{\left(\Gamma_{L1} + \Gamma_{R1}\right)\left(\Gamma_{L2} + \Gamma_{R2}\right)}\,,
\eea
which is again completely symmetric under exchange $R$ and $L$.
When one left- or right-associated tunneling rate vanishes, we obtain the current of a single-resonant level \cite{ZED09}, which is independent
of the remaining tunneling rate associated with the blocked channel: As the infinite bias limit explicitly allows double occupancy,
we do not observe Coulomb blockade in this case.
Also the infinite-bias current is well reflected in the numerical solution in Fig.~\ref{Fparcurrent_bias}.
We see that for the rate equations, the allowance of doubly occupied states simply opens more transport channels, 
which are visible by at most four steps in the current-voltage characteristics.
Interestingly, the height of these steps is independent on the level splitting, but is just determined by the
tunneling rates.
However, the (sequential tunneling) excitation spectrum of the system Hamiltonian can be probed by the
position of the current steps (sufficiently low temperatures provided).
Note that the current for $\varepsilon_1=\varepsilon_2$ provides the analytic continuation for nearly degenerate energies.
In the degenerate case the quantum master equation with coherences has to be applied as it will be shown in Sec.~\ref{BMSquantum}.
However, when additional dephasing processes lead to a decay of coherences, the rate equation approach can still be motivated.

\begin{figure}[t]
%\psfrag{PSFfinf}{$\frac{1}{2}$}
%\psfrag{PSFfcb}{$\frac{5}{9}$}
\includegraphics[width=0.47\textwidth,clip=true]{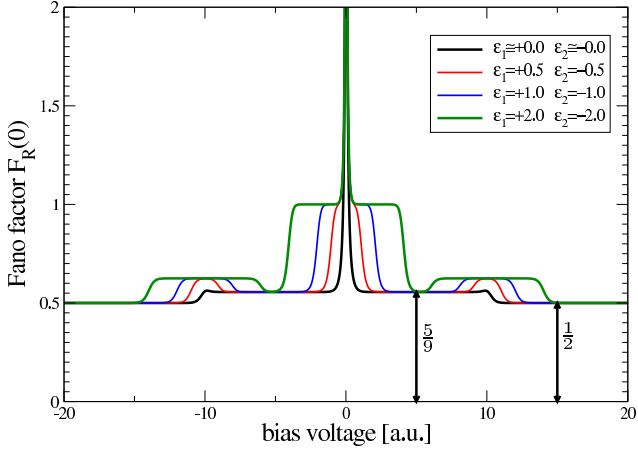}
\caption{\label{Fparfano_bias}[Color Online]
Zero frequency Fano factor (right) {\em versus} bias voltage for the parallel double dot for different single-particle energies.
The parameters correspond to those of Fig.~\ref{Fparcurrent_bias}.
The peak of the zero-frequency Fano factor at small bias voltages results from thermal (Nyquist) noise.
}
\end{figure}

For the frequency-dependent Fano factor we obtain in the Coulomb-blockade high-bias limit
\bea\label{Efano_par_cb}
F^{\rm CB}_R(\omega) = 
\frac{{\cal N}_0 + {\cal N}_1 \omega^2 + \omega^4}{{\cal D}_0 + {\cal D}_1 \omega^2 + \omega^4}
\eea
where we have chosen the abbreviations
\bea
{\cal N}_0 &=& \Gamma_{L2} \left(2 \Gamma_{L1} + \Gamma_{L2}\right) \Gamma_{R1}^2 - 2 \Gamma_{L1} \Gamma_{L2} \Gamma_{R1} \Gamma_{R2}\nn
&& + \left(\Gamma_{L1}^2 + 2 \Gamma_{L1} \Gamma_{L2} + \Gamma_{R1}^2\right) \Gamma_{R2}^2\,,\nn
{\cal N}_1 &=& \left(\Gamma_{L1} + \Gamma_{L2}\right)^2 + \Gamma_{R1}^2 + \Gamma_{R2}^2\,,\nn
{\cal D}_0 &=& \left[{\Gamma_{L2}} {\Gamma_{R1}} + \left({\Gamma_{L1}} + {\Gamma_{R1}}\right) {\Gamma_{R2}}\right]^2\,,\nn
{\cal D}_1 &=& \Big[{\Gamma_{L1}}^2 + {\Gamma_{R1}}^2 + 2 {\Gamma_{L1}} \left({\Gamma_{L2}}+ {\Gamma_{R1}}\right)\nn
&& + \left({\Gamma_{L2}} + {\Gamma_{R2}}\right)^2\Big]\,.
\eea
For zero frequency and similar left and right tunneling rates ($\Gamma_{L1}=\Gamma_{L2}$ and $\Gamma_{R1}=\Gamma_{R2}$) 
this coincides with previous results \cite{BRA06b}, see also table I(column (ii)) in Ref.~\cite{LUO07}.
As expected from the current in the CBHB regime, it can be checked easily that the zero-frequency version of the above 
Fano factor is always super-Poissonian when one of the right-associated tunneling rates vanishes (bunching due to dynamical channel blockade) and
is always sub-Poissonian when one of the left-associated tunneling rates vanishes.
Super-Poissonian Fano factors for asymmetric systems are well-known in the literature \cite{AGH06a}.
The corresponding infinite-bias Fano factor equates to
\bea\label{Efano_par_infbias}
F^\infty_R(\omega) &=& 1 -2\times\\
&& 
\frac{\frac{\Gamma_{L1}^2\Gamma_{R1}^2 \left(\Gamma_{L2}+\Gamma_{R2}\right)}{\left(\Gamma_{L1}+\Gamma_{R1}\right)^2+\omega^2}
+\frac{\left(\Gamma_{L1}+\Gamma_{R1}\right)\Gamma_{L2}^2\Gamma_{R2}^2}{\left(\Gamma_{L2}+\Gamma_{R2}\right)^2+\omega^2}}
{\Gamma_{L1}\Gamma_{L2}\left(\Gamma_{R1}+\Gamma_{R2}\right)+\left(\Gamma_{L1}+\Gamma_{L2}\right)\Gamma_{R1}\Gamma_{R2}}\,.\nonumber
\eea
For zero frequency and equal left and right tunneling rates 
($\Gamma_{L1}=\Gamma_{L2}$ and $\Gamma_{R1}=\Gamma_{R2}$) this also coincides with previous results \cite{BRA06b}, see also 
table I (column (iv)) in Ref.~\cite{LUO07}. 
From the positivity of the tunneling rates it is obvious that the above Fano factor is always sub-Poissonian, which
well matches our expectations for the infinite bias limit.
These results are well confirmed by the numerical solution, see Fig.~\ref{Fparfano_bias} for the zero-frequency Fano factor.
Note that we consider only the right-associated noise here, which only at zero frequency coincides \cite{AGU04} with the 
full frequency-resolved noise.

It follows from equations~(\ref{Efano_par_cb}) and~(\ref{Efano_par_infbias}) that for equal tunneling rates the
frequency-dependent noise in the high-bias coulomb blockade and in the infinite bias limit will essentially be featureless with a single minimum.
Numerically, we see that for equal tunneling rates this extends to the whole bias range (not shown).
However, when we assume varying tunneling rates, we may find features in the frequency-dependent noise for the CBHB regime -- 
this can be found analytically from Eqn.~(\ref{Efano_par_cb}).
These findings suggest that a feature in the frequency-dependent noise is not a unique indicator 
for coherent dynamics in the system.
This would require the evaluation of the full noise, which is the
weighted sum of noise at left and right junction and cross-correlators, see e.g. in
Ref.~\cite{AGU04}. 
This is beyond the scope of this paper.
%
%%%%%%%%%%%%%%%%%%%%%%%%%%%%%%%%%%%%%%%%%%%%%%%%%%%%%%%%%%%%%%%%%%%%%%%%%%%%%%%%%%%%

\subsection{BMS Quantum Master Equation}\label{BMSquantum}

When the master equation is considered, we assume energetic degeneracy $\epsilon_1=\epsilon_2\equiv\epsilon$ throughout.
For simplicity, we assume $\gamma_L=\sqrt{\Gamma_{L1} \Gamma_{L2}}$ and $\gamma_R=\sqrt{\Gamma_{R1} \Gamma_{R2}}$, which corresponds
to the flat band limit.
In the infinite-bias limit, we obtain the general stationary master equation current 
\bea\label{Ecurrentq_par_infbias}
I^\infty = \frac{\Gamma_{L1} \Gamma_{R1} + \Gamma_{L2} \Gamma_{R2} + 2 \sqrt{\Gamma_{L1} \Gamma_{L2} \Gamma_{R1} \Gamma_{R2}}}{\Gamma_{L1} +\Gamma_{L2} + \Gamma_{R1} + \Gamma_{R2}}
\eea
which does not necessarily coincide with the corresponding infinite bias rate equation result~(\ref{Ecurrent_par_infbias}).
For general tunneling rates we may also evaluate the frequency-dependent Fano factor analytically, which is however too lengthy to 
be reproduced here.
However, when we assume some symmetries, the expressions simplify significantly.
For example, when of the four tunneling rates, two pairs are mutually equal, one has three different
combinations, which we will label as (compare also Fig.~\ref{Fmodel_par})
\begin{itemize}
\item[I.] left-right symmetric coupling $\Gamma_{L1}=\Gamma_{R1}\equiv \Gamma_1$ with $\Gamma_{L2}=\Gamma_{R2}\equiv\Gamma_2$,
\item[II.] top-down symmetric coupling $\Gamma_{L1}=\Gamma_{L2}\equiv\Gamma_L$ with $\Gamma_{R1}=\Gamma_{R2}\equiv \Gamma_R$, and
\item[III.] antisymmetric coupling $\Gamma_{L1}=\Gamma_{R2}\equiv \Gamma_a$ with $\Gamma_{L2}=\Gamma_{R1} \equiv \Gamma_b$ (compare also \cite{DON08,SCH09})
\end{itemize}
in the following.
Note that naturally, when all tunneling rates are the same, the above three cases coincide.
In addition, we note that in all three cases, the infinite bias master equation current~(\ref{Ecurrentq_par_infbias}) and the
infinite bias rate equation current~(\ref{Ecurrent_par_infbias}) coincide.

%%%%%%%%%%%%%%%%%%%%%%%%%%%%%%%%%%%%%%%%%%%%%%%%%%%%%%%%%%%%%%%%%%%%%%%%%%%%%%%%%%%%%%%%%%%%%%%%%%%%%%%%%%%%%%%%%%%%%%%%%

\subsubsection*{Symmetric configurations: I and II}

Under the symmetry assumptions I and II, the frequency-dependent infinite bias Fano factor reduces to
\bea\label{Eparfaonoonetwothree_ib}
F_{R,I}^\infty(\omega) &=& \frac{2 \left(\Gamma_1 + \Gamma_2\right)^2 + \omega^2}{4 \left(\Gamma_1 + \Gamma_2\right)^2 + \omega^2}\,,\nn
F_{R,II}^\infty(\omega) &=& 1- \frac{8 \Gamma_L \Gamma_R}{4\left(\Gamma_L+\Gamma_R\right)^2+\omega^2}\,.
\eea
Also when the corresponding symmetries are used, these frequency-dependent Fano factors differ at finite frequency from the 
corresponding rate-equation result~(\ref{Efano_par_infbias}), the Fano factors only coincide at zero frequency.

It should be noted that for exact symmetry assumptions I and II, the null-space of the Liouville superoperator $L(0)$ becomes
two-dimensional for the complete bias range.
(Consequently, the steady state and the observables would depend on the choice of the initial state in that case.)
However, when we also consider couplings between the subspaces, we obtain an unique steady state, which is for symmetry
assumptions I and II analytically continued into Eqns.~(\ref{Eparfaonoonetwothree_ib}).
When symmetry assumptions I and II coincide (all tunneling rates equal), this becomes obvious in the basis
$\rho_{00,00}$, 
$\frac{1}{2}\left[\rho_{01,01}+\rho_{10,10}+\rho_{01,10}+\rho_{10,01}\right]$, 
$\frac{1}{2}\left[\rho_{01,01}+\rho_{10,10}-\rho_{01,10}-\rho_{10,01}\right]$,
$\rho_{11,11}$,
$\frac{1}{\sqrt{2}}\left[\rho_{01,01}-\rho_{10,10}\right]$,
$\frac{1}{\sqrt{2}}\left[\rho_{01,10}-\rho_{10,01}\right]$,
where the Liouville superoperator $L(\chi)$ decouples into three $2 \times 2$ blocks, two of which have a null-space as $\chi\to 0$.
Therefore, the above infinite bias Fano factors should be interpreted as being valid for approximate symmetries I and II.
For exactly fulfilled symmetries one will obtain two partial currents and Fano factors.
Since the finite-bias behavior is especially interesting, we will discuss these subtleties in the following at $V=U$:
In addition, we assume vanishing single-particle energies $\epsilon_1=\epsilon_2=0$ for simplicity.
When one is faced with a two-dimensional null-space of the Liouville superoperator,
one has to choose the null-space vectors orthogonal and of course normalize with respect to the trace.
Then, we obtain the currents in the two subspaces (A) and (B)
\bea\label{Ecurone}
I_{I,A}^{V=U} &=& \frac{\Gamma_1+\Gamma_2}{2} \tanh\left[\frac{\beta U}{4}\right]\,,\nn
I_{I,B}^{V=U} &=& \frac{\Gamma_1+\Gamma_2}{2} \frac{\sinh\left[\frac{\beta U}{4}\right]}{\cosh\left[\frac{3\beta U}{4}\right]}
\eea
for symmetry assumption I and similarly for symmetry assumption II
\bea\label{Ecurtwo}
I_{II,A}^{V=U} &=& 2 \frac{\Gamma_L\Gamma_R}{\Gamma_L + \Gamma_R} \tanh\left[\frac{\beta U}{4}\right]\,,\nn
I_{II,B}^{V=U} &=& 2 \frac{\Gamma_L\Gamma_R}{\Gamma_L + \Gamma_R} \frac{\sinh\left[\frac{\beta U}{4}\right]}{\cosh\left[\frac{3\beta U}{4}\right]}\,.
\eea
Naturally, for completely symmetric tunneling rates, the currents in the respective subspaces (A) and (B) of~(\ref{Ecurone}) and~(\ref{Ecurtwo})
coincide.
Regarding the noise, we obtain the frequency-dependent Fano factors for symmetry assumption I
\bea\label{Efanoone}
F_{I,A}^{V=U}(\omega) &=& \frac{2 \left(\Gamma_1+\Gamma_2\right)^2 + \omega^2}{4 \left(\Gamma_1+\Gamma_2\right)^2 + \omega^2} \coth\left[\frac{\beta U}{4}\right]\,,\nn
F_{I,B}^{V=U}(\omega) &=& \frac{{\cal N}_0 + {\cal N}_1 \omega^2}{{\cal D}_0 + {\cal D}_1 \omega^2}\,,
\eea
where we have used
${\cal N}_0 = 2 \left(\Gamma_1 + \Gamma_2\right)^2 \left(2 - 3 y + 6 y^2 - 3 y^3 + 2 y^4\right)$,
${\cal N}_1 = 1 - y + 4 y^2 - y^3 + y^4$,
${\cal D}_0 = 4 \left(\Gamma_1 + \Gamma_2\right)^2 (y-1) \left(1 + y^3\right)$,
${\cal D}_1 = (y-1) \left(1 + y^3\right)$
with
$y\equiv e^{\beta U/2}$ in the last line.
Similarly, we obtain for symmetry assumption II
\bea\label{Efanotwo}
F_{II,A}^{V=U}(\omega) &=& 
\frac{16 \Gamma_L^2 \Gamma_R y + 2 \Gamma_R \omega^2 y + \Gamma_L {\cal C}_0 \left(1 + y^2\right)}
{\Gamma_L \left[4 \left(\Gamma_L + \Gamma_R\right)^2 + \omega^2\right] \left(y^2-1\right)}\,,\nn
F_{II,B}^{V=U}(\omega) &=& \frac{{\cal N}_0 + {\cal N}_1 \omega^2}
{\Gamma_L \left[4 \left(\Gamma_L + \Gamma_R\right)^2 + \omega^2\right] (y-1) \left(1 + y^3\right)}\,,\nn
\eea
where we have used in the first equation
${\cal C}_0 = \left[4 \left(\Gamma_L^2+ \Gamma_R^2\right) + \omega^2\right]$
and in the last line
${\cal N}_0 = 4 \Gamma_L \left[2 \Gamma_L \Gamma_R {\cal P}_0 
+ \left(\Gamma_L^2+\Gamma_R^2\right) {\cal P}_1\right]$,
${\cal N}_1 = 2 \Gamma_R y^2 + \Gamma_L {\cal P}_1$
with the polynomials 
${\cal P}_0 = 1 - 2 y + 4 y^2 - 2 y^3 + y^4$, 
${\cal P}_1 = 1 - y + 2 y^2 - y^3 + y^4$,
and also as before $y\equiv e^{\beta U/2}$.
Also for the Fano factors, the expressions in equations~(\ref{Efanoone}) and~(\ref{Efanotwo}) coincide in their respective subspaces (A) and (B) 
for completely homogeneous tunneling rates.

%%%%%%%%%%%%%%%%%%%%%%%%%%%%%%%%%%%%%%%%%%%%%%%%%%%%%%%%%%%%%%%%%%%%%%%%%%%%%%%%%%%%%%%%%%%%%%%%%%%%%%%%%%%%%%%%%%%%%%%%%

\subsubsection*{Antisymmetric configuration III}

Under the assumption of an antisymmetric configuration we obtain an
unique steady state -- with the exception of completely homogeneous
couplings.

The infinite-bias Fano factor reads
\bea
F_{R,III}^\infty(\omega) &=& \frac{{\cal N}_0 + {\cal N}_1 \omega^2 + {\cal N}_2 \omega^4 + \omega^6}
{{\cal D}_0 + {\cal D}_1 \omega^2 + {\cal D}_2 \omega^4 + \omega^6}\,,
\eea
where we have used the short-hand notations
${\cal N}_0=\left(\Gamma_a - \Gamma_b\right)^4 \left(\Gamma_a^2 + \Gamma_b^2\right)$,
${\cal N}_1=\left(3 \Gamma_a + \Gamma_b\right) \left(\Gamma_a + 3 \Gamma_b\right) \left(\Gamma_a^2 + \Gamma_b^2\right)$,
${\cal N}_2=3 \left(\Gamma_a + \Gamma_b\right)^2$ and
${\cal D}_0=\left(\Gamma_a - \Gamma_b\right)^4 \left(\Gamma_a + \Gamma_b\right)^2$,
${\cal D}_1=\left(\Gamma_a - \Gamma_b\right)^4 + 2 \left(\Gamma_a + \Gamma_b\right)^2 \left(\Gamma_a^2 + 6 \Gamma_a \Gamma_b + \Gamma_b^2\right)$,
${\cal D}_2=3 \Gamma_a^2 + 14 \Gamma_a \Gamma_b + 3 \Gamma_b^2$. 
It
coincides with the corresponding rate equation
result~(\ref{Efano_par_infbias}) only at zero frequency.

\begin{figure}[t]
\includegraphics[width=0.47\textwidth,clip=true]{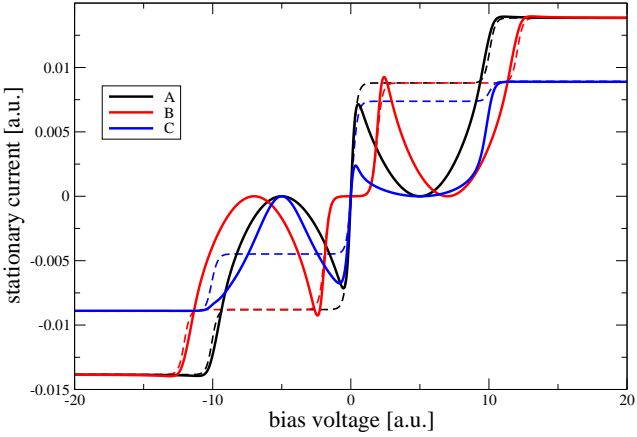}
\caption{\label{Fparcurrentq_bias}[Color Online]
Markovian stationary current predicted by the BMS quantum master equation for exact degeneracy $\epsilon_1=\epsilon_2\equiv\epsilon$.
The parameters that are the same for all curves are $U=5.0$, $\Gamma_{L1}=0.1^2$, $\Gamma_{R1}=0.15^2$, $\Gamma_{R2}=0.1^2$, $\beta=10$.
The other parameters have been varied as follows:
Case A (black): $\epsilon=0.0$, $\Gamma_{L2}=0.15^2$, 
Case B (red): $\epsilon=1.0$, $\Gamma_{L2}=0.15^2$, and 
Case C (blue): $\epsilon=0.0$, $\Gamma_{L2}=0.05^2$.
The thin dashed lines show the corresponding rate equation results.
The nearly complete suppression of the current for large $\beta U$ at $V\approx U$ is for exact degeneracy quite robust with respect
to the remaining parameters.
}
\end{figure}

\begin{figure}[t]
\includegraphics[width=0.47\textwidth,clip=true]{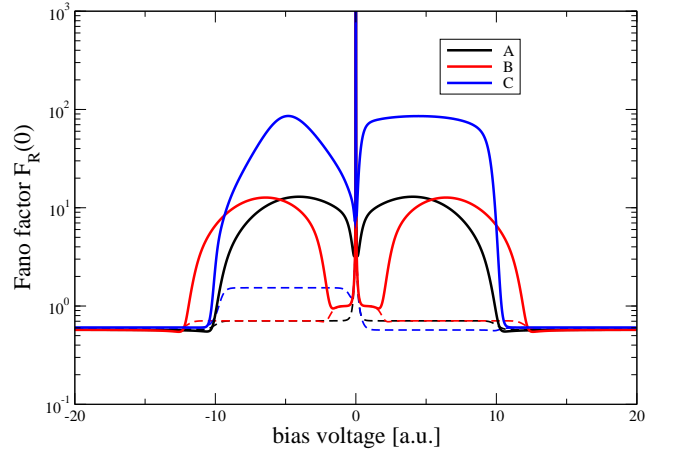}
\caption{\label{Fparfanoq_bias}[Color Online]
Zero frequency Fano factor predicted by the quantum master equation for exact degeneracy $\epsilon_1=\epsilon_2\equiv\epsilon$.
Parameters and color coding are the same as in Fig.~\ref{Fparcurrentq_bias}.
The current suppression in the quantum master equation goes along with highly 
super-Poissonian Fano factors.
}
\end{figure}

The situation becomes sophisticated when we consider the finite bias
case $V=U$ together with symmetry assumption III, where ($\Gamma_a \neq \Gamma_b$ provided) one only has one 
stationary state -- which is also the generic case for arbitrary couplings.
In this case, the Lamb-shift terms directly affect the current:
Formally, this is visible in the appearance of Digamma functions $\Psi(x)$ (resulting from the Cauchy principal value integrations, see appendix 
\ref{Alambshift}) in the stationary current.
With the replacements $y \equiv e^{\beta U/2}$, $\tilde\Gamma\equiv\Gamma_a+\Gamma_b$, $\tilde\eta\equiv\Gamma_a-\Gamma_b$ and
$\tilde \Psi(y) \equiv \Re\left[\Psi\left(\frac{1}{2} + i \frac{\ln(y)}{2\pi}\right) - \Psi\left(\frac{1}{2}-3i\frac{\ln(y)}{2\pi}\right)\right]$
we obtain with symmetry assumption III at $V=U$ and
$\epsilon_1=\epsilon_2=0$ for the current after a tedious and lengthy calculation
\begin{widetext}
\bea\label{Ecurthree}
I_{III}^{V=U} &=& 
\frac{
8 \Gamma_a \Gamma_b \tilde\Gamma y (y-1)\left(y^2+1\right) \left[\pi^2 \left(1 + y - y^2 + y^3\right) \left(3 -y +y^2+y^3\right) 
+ \left(1 + y^3\right)^2 \tilde\Psi^2(y)\right]
}
{
\pi^2 \tilde\Gamma^2 {\cal P}_1 \left(3-y+4y^2+y^4+y^5\right)^2
+ {\cal P}_1^3 {\cal P}_2^2 \left[-4 \tilde\eta^2 y {\cal P}_2 + 
\tilde\Gamma^2 \left(1 + 4 y - 2 y^2 + 4 y^3 + y^4\right)\right]\tilde\Psi^2(y)
}\,,
\eea
\end{widetext}
where we have abbreviated the polynomials
${\cal P}_1 = 1 + y$ and
${\cal P}_2 = 1 - y + y^2$. This is consistent with previous
numerical results \cite{SCH09}.
For low temperatures (large $\beta U$), the above expression behaves asymptotically as
\bea
I_{III}^{V=U}
&\stackrel{\beta U \gg 1}{\Longrightarrow}& \frac{8 \Gamma_a \Gamma_b}{\Gamma_a + \Gamma_b} e^{-\frac{\beta U}{2}}\,,
\eea
i.e., the current is exponentially suppressed for low temperatures!
The numerical solution confirms this result and shows that it goes along with a negative 
differential conductance \cite{BRA04,SCH09}, which is -- exact energetic degeneracy $\epsilon_1=\epsilon_2$ provided -- quite
robust with respect to the remaining parameters, see Fig.~\ref{Fparcurrentq_bias}.
For example, assuming non-vanishing single-particle energies $\epsilon_1=\epsilon_2\neq 0$, we see that the current suppression valley is simply shifted away
from the origin.
Even when the symmetry assumptions regarding the coupling strengths are not obeyed, the current suppression is qualitatively robust.
When we evaluate the zero-frequency Fano factor, we observe huge Fano factors in the current suppression region, 
see Fig.~\ref{Fparfanoq_bias}.
Note however, that super-Poissonian (but significantly smaller) Fano factors may also be observed in the rate equation case (due to dynamical channel blockade
as discussed before), compare also the dashed lines in Fig.~\ref{Fparcurrentq_bias}.

\begin{figure}[t]
\includegraphics[width=0.47\textwidth,clip=true]{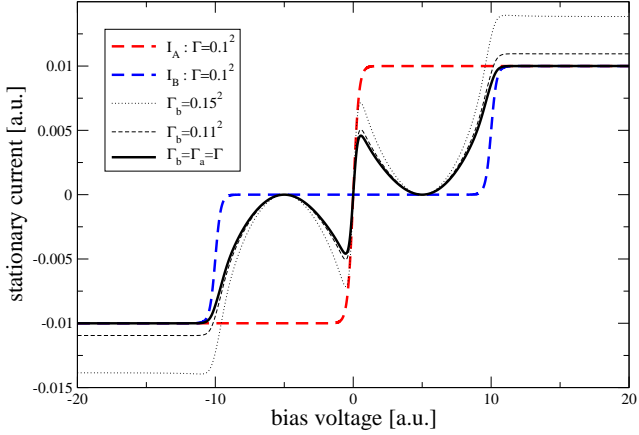}
\caption{\label{Fparcurrentpq_bias}[Color Online]
Plot of the stationary currents $I_A$ and $I_B$ for the decoupling subspaces for $\Gamma=0.1^2$ together
with the (unique) stationary current obtained for $\Gamma_a=\Gamma=0.1^2$ and different values of $\Gamma_b$ {\em versus} bias voltage.
Other parameters have been chosen as $\beta=10.0$, $U=5.0$, and $\epsilon_1=\epsilon_2=0$.
For nearly equal tunneling matrix elements, the actual current (solid line) is a convex superposition of 
the partial currents (dashed lines).
}
\end{figure}

At the point of completely homogeneous tunneling rates, we can obtain currents and Fano factors 
for the whole bias range 
\bea\label{Ecurfanoab}
I_{A} &=& \Gamma \tanh\left[\frac{\beta V}{4}\right]\,,\nn
I_{B} &=& \Gamma \frac{\sinh\left[\frac{\beta V}{2}\right]}{\cosh\left[\beta U\right]+\cosh\left[\frac{\beta V}{2}\right]}\,,\nn
F_{R,A}(0) &=& \frac{1}{2} \coth\left[\frac{\beta \abs{V}}{4}\right]\,,\nn
F_{R,B}(0) &=& \frac{\left(1 + x^2 + 2 x y^2\right) \left[2 x + \left(1 + x^2\right) y^2\right]}
{4 x \left(x + y^2\right) \left(1 + x y^2\right) \sinh\left[\frac{\beta \abs{V}}{2}\right]}\,,
\eea
where we have abbreviated $x\equiv e^{\beta V/2}$ and $y\equiv e^{\beta U/2}$ in the last line, see Fig.~\ref{Fparcurrentpq_bias}.
For $V=U$, the frequency-dependent Fano factors can be derived from 
(\ref{Efanoone}) when $\Gamma_1=\Gamma_2=\Gamma$ or from 
(\ref{Efanotwo}) when $\Gamma_L=\Gamma_R=\Gamma$.

\begin{figure}[t]
\includegraphics[width=0.47\textwidth,clip=true]{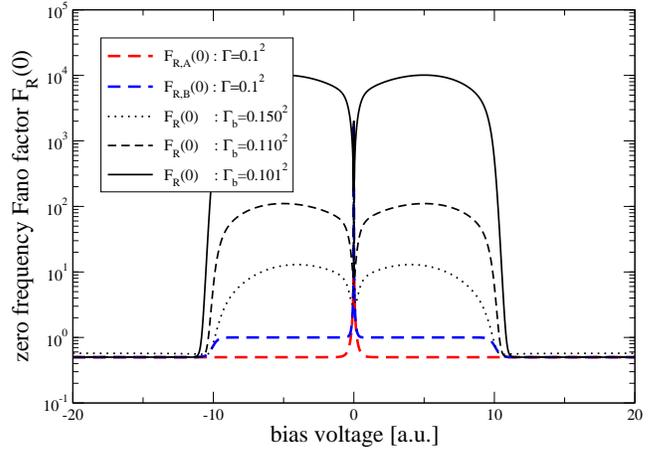}
\caption{\label{Fparfanopq_bias}[Color Online]
Logarithmic Plot of the Fano factor $F_R(0)$ {\em versus} bias voltage $-20.0\le V \le +20.0$ for different values of $\Gamma_b$.
Parameters and color coding correspond to Fig.~\ref{Fparcurrentpq_bias}.
When the partial currents do not coincide, the Fano factor diverges for $\Gamma_a=\Gamma_b$, i.e., it cannot be written as a convex
combination of the partial Fano factors (dashed red and blue lines).
}
\end{figure}

The alert reader will have noticed that in case of complete coupling symmetry 
$\Gamma_{L1}=\Gamma_{L2}=\Gamma_{R1}=\Gamma_{R2}=\Gamma$
the currents $I_{I}$ and $I_{II}$ and the Fano factors $F_{R,I}(\omega)$ and $F_{R,II}(\omega)$ 
coincide in their respective subspaces,
compare equations~(\ref{Ecurfanoab}),~(\ref{Ecurone}), and~(\ref{Ecurtwo}).
However, the antisymmetric current $I_{III}^{V=U}$ does neither converge to the current of subspace (A) nor to the
current of subspace (B) when complete symmetry is assumed, compare Eqn.~(\ref{Ecurthree}).
This demonstrates that for near degenerate tunneling rates, the actual current will rather be a superposition of the 
partial currents.
Indeed, one can find convex linear combinations of the two stationary states that reproduce the 
current $\left.I_{III}\right|_{\Gamma_a=\Gamma_b\to\Gamma}$, compare also Fig.~\ref{Fparcurrentpq_bias}.
An interesting consequence is that in this case (decoupling subspaces with different partial currents), 
it follows directly from the definition of the cumulant generating function~(\ref{Ecumgenfunc}) that 
the (zero-frequency) Fano factor must diverge (telegraph noise), which
is also seen in the numerical investigations, see Fig.~\ref{Fparfanopq_bias}.
Note that this divergence will also hold for the higher cumulants \cite{URB09}.

\begin{figure}[t]
%\psfrag{PSFheight}{$\color{orange}\propto\frac{1}{\eta^2}$}
%\psfrag{PSFwidth}{$\color{dgreen}\propto\eta^2$}
\includegraphics[width=0.47\textwidth,clip=true]{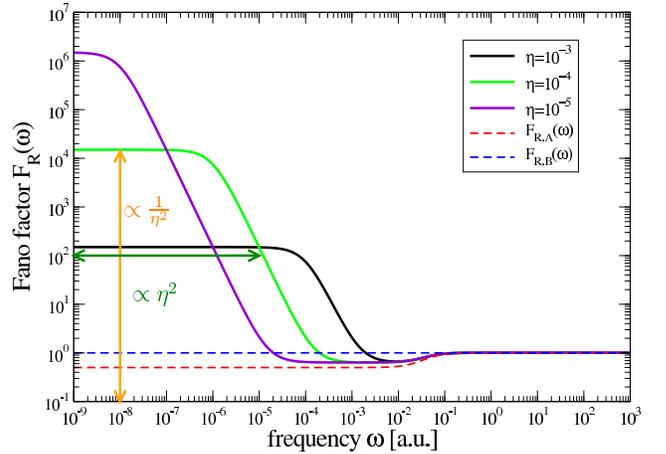}
\caption{\label{Fparnoise_sup}[Color Online]
Frequency-dependent Fano factor for symmetry assumption III at the maximum current suppression $V=U$ for 
different coupling asymmetries $\eta$.
Parameters have been chosen as $\epsilon_1=\epsilon_2=0$, $\beta=10$, $U=5$, $\Gamma_a=\Gamma-\eta$, $\Gamma_b=\Gamma+\eta$, where $\Gamma=0.1^2$.
The divergence of the actual zero-frequency Fano factor for $\eta\to 0$ manifests itself as an extremely slim peak within 
a large valley around the origin. This is similar to the spectral
Dicke effect \cite{DIC53}.
}
\end{figure}

It should be stressed that the huge Fano factors observed in Fig.~\ref{Fparfanopq_bias} around the Coulomb-blockade region 
are obtained for highly symmetric systems. For reasonable parameter
values, they are by orders of magnitude larger than the super-Poissonian Fano factors observed for
asymmetric systems \cite{AGH06a}.
In the frequency-dependent Fano factor, this divergence goes as a very narrow and tall peak within a large valley
at low frequencies, see Fig.~\ref{Fparnoise_sup}. This is similar to
the spectral Dicke effect \cite{DIC53,CON09}.
An experiment will average over a finite frequency interval determined by the inverse measurement time.
Since the area below the peak remains approximately constant (compare the scaling of height and width with respect to the asymmetry $\eta$ in Fig.~\ref{Fparnoise_sup}), 
the super-Poissonian behavior can in principle be resolved.
However, when the frequency interval of the measurement is too large, the valley may dominate the super-Poissonian peak and one may obtain
a sub-poissonian Fano factor.
Therefore, the inverse measurement time should scale with the peak width.

%%%%%%%%%%%%%%%%%%%%%%%%%%%%%%%%%%%%%%%%%%%%%%%%%%%%%%%%%%%%%%%%%%%%%%%%%%%%%%%%%%%%%%%%%%%%%%%%%%%%%%%%%%%%%%%%%%%%%%%%

\subsection{Telegraph statistics - Distribution function $P_n(t)$}

\begin{figure}[t]
\includegraphics[width=0.47\textwidth,clip=true]{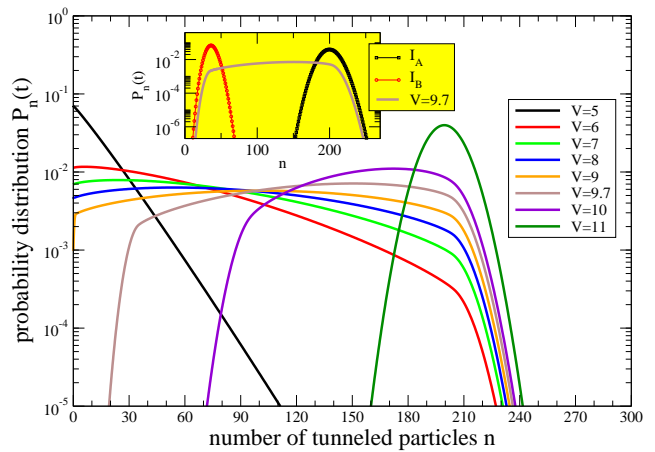}
\caption{\label{Ftelegraph_prob}[Color Online]
Normalized probability distributions $P_n(t)$ for different bias voltages {\em versus} the number of tunneled particles $n$ obtained
{\em via} the saddle-point approximation~(\ref{Esaddlepoint}).
Other parameters have been chosen as $U=5$, $\beta=10$, $t=20000$, and $\Gamma_a=\Gamma-\eta$, $\Gamma_b=\Gamma+\eta$,
where $\Gamma=0.1^2$ and $\eta=10^{-5}$.
The inset relates the probability distribution for $V=9.7$ with the corresponding partial currents (lines with symbols) 
that one would obtain in subspaces $A$ and $B$ for $\eta=0$ (cf.\cite{JOR04a}).
For bias voltages near the current suppression, the distribution becomes extremely flat -- consistent with the huge super-Poissonian 
Fano factors.
In contrast, for large bias voltages we approach the infinite bias distribution, which is characterized by a finite width.
For $V=U$, the exponential suppression of the current is manifest as an exponentially decaying distribution (straight line).
}
\end{figure}

It is quite instructive to study the impact of these results on the time-dependent probability distribution $P_n(t)$.
Given the cumulant-generating function, it is unfortunately non-trivial to obtain $P_n(t)$ {\em via} the inverse Fourier transform of equation
(\ref{Ecumgenfunc}), since that would involve an integral over a highly oscillatory function.
However, when the cumulant-generating function is highly peaked in the interval $\chi \in (-\pi, +\pi)$, we may calculate it approximately
{\em via} the saddle-point approximation (also termed stationary phase approximation, see e.g. Refs.~\cite{BAG03a,KIE05a,SCH06})
\bea\label{Esaddlepoint}
P_n(t) &=& \frac{1}{2\pi} \int\limits_{-\pi}^{+\pi} e^{S(\chi,t) - i n \chi} d\chi\nn
&\approx& \frac{e^{S(\chi^*,t)-i n \chi^*}}{2\pi} \int\limits_{-\pi}^{+\pi} e^{S''(\chi^*,t) (\chi-\chi^*)^2/2} d\chi\,,
\eea
where we see that the remaining integral in the above equation just corresponds to a normalization of the distribution, since it does not depend on $n$.
The position $\chi^*$ of the integrand peak is determined by the equation
\bea\label{Epos_saddle}
\left.\partial_{\chi} S(\chi,t)\right|_{\chi=\chi^*} = i n\,,
\eea
which only admits purely imaginary solutions for $\chi^*$.
Since Eqn.~(\ref{Epos_saddle}) is rather demanding to solve numerically, we have computed $n$ 
from Eqn.~(\ref{Epos_saddle}) and the corresponding distribution
$P_n(t) \propto \exp\left\{S(\chi^*,t)-i n \chi^*\right\}$ parametrically as a function of all imaginary $\chi^*$.
Normalization was then performed afterwards, see Fig.~\ref{Ftelegraph_prob}.
Instead of obtaining a bimodal distribution as might be naively expected from the existence of two different currents in the
subspaces A and B, we see that the actual probability distribution is
unimodal \cite{JOR04a}.
The mean of the distributions -- divided by the snapshot time $t$ -- yields the current at the corresponding bias voltage, which
is well consistent with the current-voltage characteristics in Fig.~\ref{Fparcurrentpq_bias}.
The huge zero-frequency Fano factors for bias voltages below the Coulomb interaction strength -- compare Fig.~\ref{Fparfanopq_bias} -- 
are reflected in extremely large widths for the corresponding bias range.
Note however that Fig.~\ref{Ftelegraph_prob} shows a finite time snapshot which does not capture the large time limit 
of the zero-frequency Fano factor.

%%%%%%%%%%%%%%%%%%%%%%%%%%%%%%%%%%%%%%%%%%%%%%%%%%%%%%%%%%%%%%%%%%%%%%%%%%%%%%%%%%%%

\subsection{Coarse-Graining Results}

Although the BMS currents do not display completely unphysical behavior, it is disturbing that
the BMS Liouvillian behaves discontinuously as a function of its parameters, which also transfers
to observables such as the current.

\begin{figure}[t]
\includegraphics[width=0.47\textwidth,clip=true]{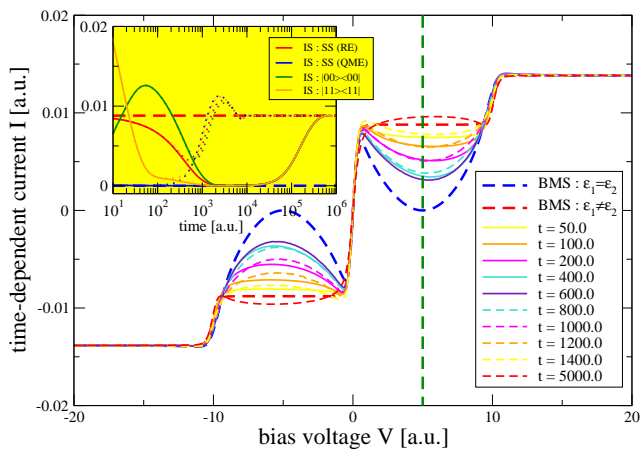}
\caption{\label{Fparcurrent_tdep}[Color Online]
Plot of the time-dependent DCG current {\em versus} bias voltage for different times.
The other parameters were chosen as $\epsilon_1=+0.0005$, $\epsilon_2=-0.0005$, $U=5.0$, $\beta=10.0$,
$\Gamma_{L1}=\Gamma_{R2}=0.1^2$, and $\Gamma_{L2}=\Gamma_{R1}=0.15^2$.
In the plot, the initial condition has been chosen as the stationary density matrix of the 
BMS rate equation, whereas the inset shows the actual time-dependence of the current at $V=U$ (compare the vertical green dashed line) 
for different initial conditions:
the steady state density matrix of the BMS rate equation (SS-RE), the stationary density matrix of the BMS master equation (SS-QME),
the empty system, and the doubly occupied system.
For times smaller than the inverse level splitting $\Delta E^{-1} = 1000$ (thin solid lines), we observe a transient crossover from the 
BMS rate equation current (bold dashed red line) towards the BMS master equation current (bold dashed blue line).
When the time approaches and exceeds the inverse level splitting, this trend is reversed (thin dashed lines), 
until one recovers the initial dynamics.
The inset shows that this behavior is qualitatively independent on the initial state (IS) and 
that for significantly smaller splittings $\delta^{-1}=10^5$ (solid lines) in comparison to $\delta^{-1}=10^3$ (dotted lines)
the two transient processes do well separate.
}
\end{figure}

One may argue that in a realistic setting, there will always exist lifted degeneracies, such that 
for the stationary state the result of the rate equation is relevant.
In this case, the DCG approach will finally approach the BMS rate equation current, but the BMS master
equation current will appear as a metastable state -- regardless of the initial state.
These expectations are also found in the numerical solution, see Fig.~\ref{Fparcurrent_tdep}.
For small times $t=\tau$, the transient behavior resulting from the initial state is still found in the stationary current 
(solid lines in Fig.~\ref{Fparcurrent_tdep}), but the relaxation drags the current towards the
master equation result.
For times larger than the inverse level splitting, we observe a relaxation back towards the rate equation result 
(which had also been chosen as the initial state in Fig.~\ref{Fparcurrent_tdep}).
However, as the timescales of relaxation ($\Gamma^{-1}=100$) and the
inverse level splitting ($\delta^{-1}=1000$) do
not completely separate, we do not observe a complete decay into the master equation result.
The inset in Fig.~\ref{Fparcurrent_tdep} shows that this behavior is independent of the particular initial state and
that for separating time scales a complete decay into the metastable state is observed.

\begin{figure}[t]
\includegraphics[width=0.47\textwidth,clip=true]{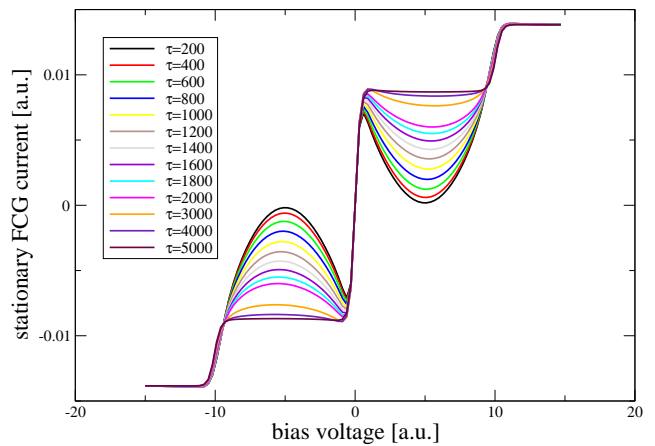}
\caption{\label{FparcurrentFCG_bias}[Color Online]
Plot of the stationary current {\em versus} bias voltage for different coarse-graining times $\tau$.
For small coarse-graining times, we recover the qualitative behavior of the BMS master equation, whereas
for large coarse-graining times, the BMS rate equation result is approached.
The other parameters have been chosen as $\epsilon_1=+0.0005$, $\epsilon_2=-0.0005$, $U=5.0$, 
$\Gamma_{L1}=\Gamma_{R2}=0.1^2$, $\Gamma_{L2}=\Gamma_{R1}=0.15^5$, and $\beta=10.0$.
}
\end{figure}

Inspired by the success of finite maximum coarse-graining times~(\ref{Ecoarsegrainingmax}) for the serial model in Sec.~\ref{SSScoarse_graining_results}, 
we have also calculated the stationary current as a function of the coarse graining time.
We find that the finite coarse-graining (FCG) approach also predicts the qualitative effects of
the BMS master equation in the stationary state, see Fig.~\ref{FparcurrentFCG_bias}.
By varying the coarse-graining time we actually observe a smooth crossover from the quantum master equation results
(small coarse-graining times) towards the rate equation results (large coarse-graining times).
This directly demonstrates that the singular coupling limit is not the only master equation method 
yielding Lindblad type master equations that smoothly interpolates between rate equation and quantum master
equation results \cite{SCH09}.
Note that alternatively, we could have taken the coarse-graining time as constant and modified the level splitting $\epsilon_1-\epsilon_2$, 
compare also Fig.~\ref{FsercurrentFCG_bias}.

Provided that the measurement time exceeds the (non-vanishing) inverse
level splitting, experiments should be able to back up either the
FCG results (e.g., when a negative differential conductance is
measured) or the DCG results (steplike current voltage characteristics)
just by measuring the current in a fixed time interval.

%%%%%%%%%%%%%%%%%%%%%%%%%%%%%%%%%%%%%%%%%%%%%%%%%%%%%%%%%%%%%%%%%%%%%%%%%%%%%%%%%%%%
%%%%%%%%%%%%%%%%%%%%%%%%%%%%%%%%%%%%%%%%%%%%%%%%%%%%%%%%%%%%%%%%%%%%%%%%%%%%%%%%%%%%
%%%%%%%%%%%%%%%%%%%%%%%%%%%%%%%%%%%%%%%%%%%%%%%%%%%%%%%%%%%%%%%%%%%%%%%%%%%%%%%%%%%%
%%%%%%%%%%%%%%%%%%%%%%%%%%%%%%%%%%%%%%%%%%%%%%%%%%%%%%%%%%%%%%%%%%%%%%%%%%%%%%%%%%%%
\section{Summary}\label{Ssummary}
%%%%%%%%%%%%%%%%%%%%%%%%%%%%%%%%%%%%%%%%%%%%%%%%%%%%%%%%%%%%%%%%%%%%%%%%%%%%%%%%%%%%
%%%%%%%%%%%%%%%%%%%%%%%%%%%%%%%%%%%%%%%%%%%%%%%%%%%%%%%%%%%%%%%%%%%%%%%%%%%%%%%%%%%%
%%%%%%%%%%%%%%%%%%%%%%%%%%%%%%%%%%%%%%%%%%%%%%%%%%%%%%%%%%%%%%%%%%%%%%%%%%%%%%%%%%%%
%%%%%%%%%%%%%%%%%%%%%%%%%%%%%%%%%%%%%%%%%%%%%%%%%%%%%%%%%%%%%%%%%%%%%%%%%%%%%%%%%%%%

We have studied single-electron transport through serial and parallel configurations of two interacting levels.
We considered the weak-coupling limit with arbitrary Coulomb interaction strengths and
bias voltages. 
Our calculations in the energy eigenbasis 
included couplings between populations and coherences, which are
mediated by the Lamb-shift.

We have derived the $n$-resolved quantum master equation by virtue of an
auxiliary detector at one junction.
Obtaining the Liouville superoperator by coarse-graining generally
leads to Lindblad-type master equations.
The BMS approximation naturally arises in the limit of infinite coarse-graining
times. As a remarkable property, we find that it leads to equilibration of both
temperatures and chemical potentials between system and reservoir for
zero bias.
Unfortunately, the BMS Liouvillian depends discontinuously on the parameters of the
system Hamiltonian:
For a degenerate spectrum, coherences couple to the populations in the
system energy eigenbasis, whereas for a non-degenerate system
spectrum, coherences and populations evolve independently.
We demonstrate that for finite coarse-graining times this is not the
case.

In the serial configuration, we observe a steplike increase of the current {\em versus} bias
voltage caused by additional transport channels beyond the
Coulomb-blockade regime.
However, the neglect of coherences is only valid for large values of
the interdot tunnel coupling:
For vanishing interdot tunneling, the BMS current remains finite,
which is clearly unphysical.
With coherences (finite coarse-graining times), the stationary current
vanishes quadratically for small interdot tunneling in agreement with
exact results.
In the Coulomb-blockade regime, the noise spectrum at one junction may
display additional minima, and the zero-frequency Fano factor
indicates bunching for highly asymmetric lead couplings.

In the parallel case, one may also obtain super-Poissonian Fano
factors for the rate equation case (neglect of coherences) for
asymmetric lead couplings due to dynamical channel blockade.
In contrast, the quantum master equation predicts giant Fano factors
for symmetric configurations due to interference effects.
This goes along with exponential current suppression for sufficiently
low temperatures in the Coulomb-blockade regime.
For perfectly symmetric lead couplings, the quantum master equation dynamics decouples into
two independent subspaces of physical relevance, which bears strong similarities to classical
telegraph signals.
In the frequency-dependent noise spectrum, this zero-frequency
divergence appears as a $\delta$-like peak, similar to
the spectral Dicke effect.
The time-dependent DCG current displays intriguing non-Markovian features
such as the temporary dwell in a  metastable state before eventually
relaxing into the BMS steady state.
However, the final decay will only take place when the maximum
coarse-graining time is larger than the inverse splitting between the
singly-charged eigenstates.

Future research should clarify the question which maximum optimal coarse-graining time
should be chosen in order to obtain the best stationary 
observables. I.e., so far we only have discussed the effect of finite coarse-graining times qualitatively.
A direct comparison of the FCG approach within exactly solvable models might give a quantitative estimate
on the maximum coarse graining time $\tau_\textrm{max}$ and the evolution $\tau (t)$ [see Eq.~(\ref{Ecoarsegrainingmax})].

%%%%%%%%%%%%%%%%%%%%%%%%%%%%%%%%%%%%%%%%%%%%%%%%%%%%%%%%%%%%%%%%%%%%%%%%%%%%%%%%%%%%
%%%%%%%%%%%%%%%%%%%%%%%%%%%%%%%%%%%%%%%%%%%%%%%%%%%%%%%%%%%%%%%%%%%%%%%%%%%%%%%%%%%%
%%%%%%%%%%%%%%%%%%%%%%%%%%%%%%%%%%%%%%%%%%%%%%%%%%%%%%%%%%%%%%%%%%%%%%%%%%%%%%%%%%%%
%%%%%%%%%%%%%%%%%%%%%%%%%%%%%%%%%%%%%%%%%%%%%%%%%%%%%%%%%%%%%%%%%%%%%%%%%%%%%%%%%%%%
\begin{acknowledgments}

Constructive discussions with M. Schultz, J. K\"onig, D. Urban, and A. Wacker are gratefully acknowledged.
The authors are especially indebted to C. Emary for running his algorithm on the parallel 
quantum dot model.
This work was supported financially by the Deutsche Forschungsgemeinschaft under Grant No. BR 1528/5-1.

\end{acknowledgments}

%%%%%%%%%%%%%%%%%%%%%%%%%%%%%%%%%%%%%%%%%%%%%%%%%%%%%%%%%%%%%%%%%%%%%%%%%%%%%%%%%%%%
%%%%%%%%%%%%%%%%%%%%%%%%%%%%%%%%%%%%%%%%%%%%%%%%%%%%%%%%%%%%%%%%%%%%%%%%%%%%%%%%%%%%
%%%%%%%%%%%%%%%%%%%%%%%%%%%%%%%%%%%%%%%%%%%%%%%%%%%%%%%%%%%%%%%%%%%%%%%%%%%%%%%%%%%%
%%%%%%%%%%%%%%%%%%%%%%%%%%%%%%%%%%%%%%%%%%%%%%%%%%%%%%%%%%%%%%%%%%%%%%%%%%%%%%%%%%%%

%%%%%%%%%%%%%%%%%%%%%%%%%%%%%%%%%%%%%%%%%%%%%%%%%%%%%%%%%%%%%%%%%%%%%%%%%%%%%%%%%%%%
%%%%%%%%%%%%%%%%%%%%%%%%%%%%%%%%%%%%%%%%%%%%%%%%%%%%%%%%%%%%%%%%%%%%%%%%%%%%%%%%%%%%
%%%%%%%%%%%%%%%%%%%%%%%%%%%%%%%%%%%%%%%%%%%%%%%%%%%%%%%%%%%%%%%%%%%%%%%%%%%%%%%%%%%%
%%%%%%%%%%%%%%%%%%%%%%%%%%%%%%%%%%%%%%%%%%%%%%%%%%%%%%%%%%%%%%%%%%%%%%%%%%%%%%%%%%%%
\begin{appendix}

%%%%%%%%%%%%%%%%%%%%%%%%%%%%%%%%%%%%%%%%%%%%%%%%%%%%%%%%%%%%%%%%%%%%%%%%%%%%%%%%%%%%
%%%%%%%%%%%%%%%%%%%%%%%%%%%%%%%%%%%%%%%%%%%%%%%%%%%%%%%%%%%%%%%%%%%%%%%%%%%%%%%%%%%%
%%%%%%%%%%%%%%%%%%%%%%%%%%%%%%%%%%%%%%%%%%%%%%%%%%%%%%%%%%%%%%%%%%%%%%%%%%%%%%%%%%%%
%%%%%%%%%%%%%%%%%%%%%%%%%%%%%%%%%%%%%%%%%%%%%%%%%%%%%%%%%%%%%%%%%%%%%%%%%%%%%%%%%%%%

\section{Serial Coarse-Graining Liouvillian}\label{Aliouville_ser}

By transforming the bath coupling operators in equations~(\ref{Ecoup_ser}) to the interaction picture we obtain
with using the tunneling rates~(\ref{Etunnelser}) the bath correlation functions.
The Fourier transforms~(\ref{Efourier}) of the nonvanishing bath correlation functions are simply given by
\bea\label{Eserftcorrfunc}
\gamma_{12}(\omega) &=& \Gamma_L(\omega) f_L(\omega)\,,\nn
\gamma_{21}(\omega) &=& \Gamma_L(-\omega) \left[1-f_L(-\omega)\right]\,,\nn
\gamma_{34}(\omega) &=& \Gamma_R(\omega) f_R(\omega)\,,\nn
\gamma_{43}(\omega) &=& \Gamma_R(-\omega) \left[1-f_R(-\omega)\right]\,.
\eea
When we make use of the relations 
\mbox{$\Theta(t_1-t_2)=\frac{1}{2}\left[1+ {\rm sgn}(t_1-t_2)\right]$} and also
\mbox{$\Theta(t_2-t_1)=\frac{1}{2}\left[1- {\rm sgn}(t_1-t_2)\right]$}, we can insert the even and odd 
Fourier transforms~(\ref{Efourier}) into the coarse-graining Liouvillian~(\ref{Edcg_liouville}).
The Liouvillian separates into a dissipative ($\propto \gamma_{ij}(\omega)$ and a unitary (Lamb-shift) part
($\propto \sigma_{ij}(\omega)$),
\begin{widetext}
\bea\label{Eliouville_ser}
L^\tau [\f{\rho}] 
&=& \frac{1}{2\pi} \int\limits_{-\infty}^{+\infty} d\omega \frac{1}{\tau} \int\limits_0^\tau dt_1 dt_2 e^{+i \omega(t_1-t_2)}\Big\{\nn
&&+\gamma_{12}(\omega)\left[+\f{A_2}(t_2) \f{\rho} \f{A_2^\dagger}(t_1)-\frac{1}{2} \left\{ \f{A_2^\dagger}(t_1) \f{A_2}(t_2), \f{\rho}\right\}\right]
-i \left[\frac{\sigma_{12}(\omega)}{2i} \f{A_2^\dagger}(t_1) \f{A_2}(t_2), \f{\rho} \right]\nn
&&+\gamma_{21}(\omega)\left[+\f{A_1}(t_2) \f{\rho} \f{A_1^\dagger}(t_1)-\frac{1}{2} \left\{ \f{A_1^\dagger}(t_1) \f{A_1}(t_2), \f{\rho}\right\}\right]
-i \left[\frac{\sigma_{21}(\omega)}{2i} \f{A_1^\dagger}(t_1) \f{A_1}(t_2), \f{\rho} \right]\nn
&&+\gamma_{34}(\omega)\left[+\f{A_4}(t_2) \f{\rho} \f{A_4^\dagger}(t_1)-\frac{1}{2} \left\{ \f{A_4^\dagger}(t_1) \f{A_4}(t_2), \f{\rho}\right\}\right]
-i \left[\frac{\sigma_{34}(\omega)}{2i} \f{A_4^\dagger}(t_1) \f{A_4}(t_2), \f{\rho} \right]\nn
&&+\gamma_{43}(\omega)\left[+\f{A_3}(t_2) \f{\rho} \f{A_3^\dagger}(t_1)-\frac{1}{2} \left\{ \f{A_3^\dagger}(t_1) \f{A_3}(t_2), \f{\rho}\right\}\right]
-i \left[\frac{\sigma_{43}(\omega)}{2i} \f{A_3^\dagger}(t_1) \f{A_3}(t_2), \f{\rho} \right]
\Big\}\,.
\eea
\end{widetext}
In appendix~\ref{Alambshift} we demonstrate how for Lorentzian-shaped bands $\Gamma(\omega)$ the odd Fourier transforms $\sigma_{ij}(\omega)$ can be extracted
analytically from the even Fourier transforms $\gamma_{ij}(\omega)$ given in Eqn.~(\ref{Eserftcorrfunc}).
The time-dependence of the system operators arises from the interaction picture, it can always be written as a sum over oscillatory
terms (eigenoperator decomposition \cite{BRE02,SCH08}).
Therefore, the time integrations in Eqn.~(\ref{Eliouville_ser}) can be performed analytically, e.g. as 
\bea
\int\limits_0^\tau e^{i \left(\omega-\omega_a\right) t_1} dt_1 
= \tau e^{i \left(\omega-\omega_a\right) \tau/2} {\rm sinc}\left[\left(\omega-\omega_a\right) \frac{\tau}{2}\right]
\eea
and similarly for the other integral, where the band filter function ${\rm sinc}(x) \equiv \frac{\sin(x)}{x}$ 
has been introduced.
Due to the two time integrations, products of two band filter functions arise, and in the large $\tau$ limit
we may use for discrete $\omega_a,\omega_b$ the identity \cite{SCH08}
\bea\label{Esincidentity}
\Delta(\omega_a, \omega_b) &\equiv& \lim_{\tau\to\infty} \frac{\tau}{2\pi} {\rm sinc}\left[(\omega-\omega_a)\frac{\tau}{2}\right] 
{\rm sinc}\left[(\omega-\omega_b)\frac{\tau}{2}\right]\nn
&=& \delta_{\omega_a,\omega_b} \delta(\omega-\omega_a)\,,
\eea
where $\delta_{\omega_a,\omega_b}$ denotes the Kronecker symbol and $\delta(\omega-\omega_a)$ the Dirac $\delta$ distribution.
Thus, in the limit $\tau\to\infty$ also the frequency integral in~(\ref{Eliouville_ser}) collapses and we obtain
the BMS Liouvillian.

%%%%%%%%%%%%%%%%%%%%%%%%%%%%%%%%%%%%%%%%%%%%%%%%%%%%%%%%%%%%%%%%%%%%%%%%%%%%%%%%%%%%
%%%%%%%%%%%%%%%%%%%%%%%%%%%%%%%%%%%%%%%%%%%%%%%%%%%%%%%%%%%%%%%%%%%%%%%%%%%%%%%%%%%%

\section{Parallel Coarse-Graining Liouvillian}\label{Aliouville_par}

To obtain the interaction picture, it is advantageous \cite{BRE02} to expand the system coupling operators
in~(\ref{Ecoup_par}) into eigenoperators of the system Hamiltonian, as for example
$\f{d_1}(t) = e^{+i \HS t} \left( d_1 d_2^\dagger d_2 + d_1 d_2 d_2^\dagger \right) e^{-i\HS t} 
= e^{-i(\epsilon_1+U)t} d_1 d_2^\dagger d_2 + e^{-i\epsilon_1 t} d_1 d_2 d_2^\dagger$
and similarly for the other operators.
With the tunneling rates~(\ref{Etunnelpar}) we obtain for the Fourier-transforms~(\ref{Efourier}) of the non-vanishing
bath correlation functions the result
\bea\label{Eparftcorrfunc}
\gamma_{12}(\omega) &=& \Gamma_{R1}(\omega) f_R(\omega)\,,\qquad
\gamma_{14}(\omega) = \gamma_R(\omega) f_R(\omega)\,,\nn
\gamma_{21}(\omega) &=& \Gamma_{R1}(-\omega) \left[1-f_R(-\omega)\right]\,,\nn
\gamma_{23}(\omega) &=& \gamma_R^*(-\omega) \left[1-f_R(-\omega)\right]\,,\nn
\gamma_{32}(\omega) &=& \gamma_R^*(\omega) f_R(\omega)\,,\qquad
\gamma_{34}(\omega) = \Gamma_{R2}(\omega) f_R(\omega)\,,\nn
\gamma_{41}(\omega) &=& \gamma_R(-\omega) \left[1-f_R(-\omega)\right]\,,\nn
\gamma_{43}(\omega) &=& \Gamma_{R2}(-\omega) \left[1-f_R(-\omega)\right]\,,\nn
\gamma_{56}(\omega) &=& \Gamma_{L1}(\omega) f_L(\omega)\,,\qquad
\gamma_{58}(\omega) = \gamma_L(\omega) f_L(\omega)\,,\nn
\gamma_{65}(\omega) &=& \Gamma_{L1}(-\omega) \left[1-f_L(-\omega)\right]\,,\nn
\gamma_{67}(\omega) &=& \gamma_L^*(-\omega) \left[1-f_L(-\omega)\right]\,,\nn
\gamma_{76}(\omega) &=& \gamma_L^*(\omega) f_L(\omega)\,,\qquad
\gamma_{78}(\omega) = \Gamma_{L2}(\omega) f_L(\omega)\,,\nn
\gamma_{85}(\omega) &=& \gamma_L(-\omega) \left[1-f_L(-\omega)\right]\,,\nn
\gamma_{87}(\omega) &=& \Gamma_{L2}(-\omega) \left[1-f_L(-\omega)\right]\,.
\eea
As these Fourier transforms directly determine the dissipative part of the Liouvillian, this
also demonstrates that one cannot always identify the latter with the real part of the Liouvillian (the tunneling rates 
$\gamma_R(\omega)$ and $\gamma_L(\omega)$ may be chosen complex-valued).
Making use of 
\mbox{$\Theta(t_1-t_2) = \frac{1}{2}\left[1+{\rm sgn}(t_1-t_2)\right]$}
and 
\mbox{$\Theta(t_1-t_2) = \frac{1}{2}\left[1-{\rm sgn}(t_1-t_2)\right]$}
we can insert both even and odd Fourier transforms~(\ref{Efourier}) 
into the coarse-graining Liouvillian~(\ref{Edcg_liouville}).
This leads to
\begin{widetext}
\bea
L^\tau [\f{\rho}] &=& \frac{1}{2\pi} \int\limits_{-\infty}^{+\infty} d\omega \frac{1}{\tau} \int\limits_0^\tau dt_1 dt_2 e^{+i \omega (t_1-t_2)}\Big\{\nn
&&+\gamma_{12}(\omega)\left[ +\f{A_2}(t_2) \f{\rho} \f{A_2^\dagger}(t_1) - \frac{1}{2}\left\{ \f{A_2^\dagger}(t_1) \f{A_2}(t_2), \f{\rho} \right\}\right]
-i \left[ \frac{\sigma_{12}(\omega)}{2i} \f{A_2^\dagger}(t_1) \f{A_2}(t_2), \f{\rho} \right]\nn
&&+\gamma_{14}(\omega)\left[ +\f{A_4}(t_2) \f{\rho} \f{A_2^\dagger}(t_1) - \frac{1}{2}\left\{ \f{A_2^\dagger}(t_1) \f{A_4}(t_2), \f{\rho} \right\}\right]
-i \left[ \frac{\sigma_{14}(\omega)}{2i} \f{A_2^\dagger}(t_1) \f{A_4}(t_2), \f{\rho} \right]\nn
&&+\gamma_{21}(\omega)\left[ +\f{A_1}(t_2) \f{\rho} \f{A_1^\dagger}(t_1) - \frac{1}{2}\left\{ \f{A_1^\dagger}(t_1) \f{A_1}(t_2), \f{\rho} \right\}\right]
-i \left[ \frac{\sigma_{21}(\omega)}{2i} \f{A_1^\dagger}(t_1) \f{A_1}(t_2), \f{\rho} \right]\nn
&&+\gamma_{23}(\omega)\left[ +\f{A_3}(t_2) \f{\rho} \f{A_1^\dagger}(t_1) - \frac{1}{2}\left\{ \f{A_1^\dagger}(t_1) \f{A_3}(t_2), \f{\rho} \right\}\right]
-i \left[ \frac{\sigma_{23}(\omega)}{2i} \f{A_1^\dagger}(t_1) \f{A_3}(t_2), \f{\rho} \right]\nn
&&+\gamma_{32}(\omega)\left[ +\f{A_2}(t_2) \f{\rho} \f{A_4^\dagger}(t_1) - \frac{1}{2}\left\{ \f{A_4^\dagger}(t_1) \f{A_2}(t_2), \f{\rho} \right\}\right]
-i \left[ \frac{\sigma_{32}(\omega)}{2i} \f{A_4^\dagger}(t_1) \f{A_2}(t_2), \f{\rho} \right]\nn
&&+\gamma_{34}(\omega)\left[ +\f{A_4}(t_2) \f{\rho} \f{A_4^\dagger}(t_1) - \frac{1}{2}\left\{ \f{A_4^\dagger}(t_1) \f{A_4}(t_2), \f{\rho} \right\}\right]
-i \left[ \frac{\sigma_{34}(\omega)}{2i} \f{A_4^\dagger}(t_1) \f{A_4}(t_2), \f{\rho} \right]\nn
&&+\gamma_{41}(\omega)\left[ +\f{A_1}(t_2) \f{\rho} \f{A_3^\dagger}(t_1) - \frac{1}{2}\left\{ \f{A_3^\dagger}(t_1) \f{A_1}(t_2), \f{\rho} \right\}\right]
-i \left[ \frac{\sigma_{41}(\omega)}{2i} \f{A_3^\dagger}(t_1) \f{A_1}(t_2), \f{\rho} \right]\nn
&&+\gamma_{43}(\omega)\left[ +\f{A_3}(t_2) \f{\rho} \f{A_3^\dagger}(t_1) - \frac{1}{2}\left\{ \f{A_3^\dagger}(t_1) \f{A_3}(t_2), \f{\rho} \right\}\right]
-i \left[ \frac{\sigma_{43}(\omega)}{2i} \f{A_3^\dagger}(t_1) \f{A_3}(t_2), \f{\rho} \right]\nn
&&+\gamma_{56}(\omega)\left[ +\f{A_6}(t_2) \f{\rho} \f{A_6^\dagger}(t_1) - \frac{1}{2}\left\{ \f{A_6^\dagger}(t_1) \f{A_6}(t_2), \f{\rho} \right\}\right]
-i \left[ \frac{\sigma_{56}(\omega)}{2i} \f{A_6^\dagger}(t_1) \f{A_6}(t_2), \f{\rho} \right]\nn
&&+\gamma_{58}(\omega)\left[ +\f{A_8}(t_2) \f{\rho} \f{A_6^\dagger}(t_1) - \frac{1}{2}\left\{ \f{A_6^\dagger}(t_1) \f{A_8}(t_2), \f{\rho} \right\}\right]
-i \left[ \frac{\sigma_{58}(\omega)}{2i} \f{A_6^\dagger}(t_1) \f{A_8}(t_2), \f{\rho} \right]\nn
&&+\gamma_{65}(\omega)\left[ +\f{A_5}(t_2) \f{\rho} \f{A_5^\dagger}(t_1) - \frac{1}{2}\left\{ \f{A_5^\dagger}(t_1) \f{A_5}(t_2), \f{\rho} \right\}\right]
-i \left[ \frac{\sigma_{65}(\omega)}{2i} \f{A_5^\dagger}(t_1) \f{A_5}(t_2), \f{\rho} \right]\nn
&&+\gamma_{67}(\omega)\left[ +\f{A_7}(t_2) \f{\rho} \f{A_5^\dagger}(t_1) - \frac{1}{2}\left\{ \f{A_5^\dagger}(t_1) \f{A_7}(t_2), \f{\rho} \right\}\right]
-i \left[ \frac{\sigma_{67}(\omega)}{2i} \f{A_5^\dagger}(t_1) \f{A_7}(t_2), \f{\rho} \right]\nn
&&+\gamma_{76}(\omega)\left[ +\f{A_6}(t_2) \f{\rho} \f{A_8^\dagger}(t_1) - \frac{1}{2}\left\{ \f{A_8^\dagger}(t_1) \f{A_6}(t_2), \f{\rho} \right\}\right]
-i \left[ \frac{\sigma_{76}(\omega)}{2i} \f{A_8^\dagger}(t_1) \f{A_6}(t_2), \f{\rho} \right]\nn
&&+\gamma_{78}(\omega)\left[ +\f{A_8}(t_2) \f{\rho} \f{A_8^\dagger}(t_1) - \frac{1}{2}\left\{ \f{A_8^\dagger}(t_1) \f{A_8}(t_2), \f{\rho} \right\}\right]
-i \left[ \frac{\sigma_{78}(\omega)}{2i} \f{A_8^\dagger}(t_1) \f{A_8}(t_2), \f{\rho} \right]\nn
&&+\gamma_{85}(\omega)\left[ +\f{A_5}(t_2) \f{\rho} \f{A_7^\dagger}(t_1) - \frac{1}{2}\left\{ \f{A_7^\dagger}(t_1) \f{A_5}(t_2), \f{\rho} \right\}\right]
-i \left[ \frac{\sigma_{85}(\omega)}{2i} \f{A_7^\dagger}(t_1) \f{A_5}(t_2), \f{\rho} \right]\nn
&&+\gamma_{87}(\omega)\left[ +\f{A_7}(t_2) \f{\rho} \f{A_7^\dagger}(t_1) - \frac{1}{2}\left\{ \f{A_7^\dagger}(t_1) \f{A_7}(t_2), \f{\rho} \right\}\right]
-i \left[ \frac{\sigma_{87}(\omega)}{2i} \f{A_7^\dagger}(t_1) \f{A_7}(t_2), \f{\rho} \right]
\Big\}\,.
\eea
\end{widetext}
We demonstrate in appendix~\ref{Alambshift} how for Lorentzian-shaped bands $\Gamma(\omega)$ the odd Fourier transforms $\sigma_{ij}(\omega)$ can be extracted
analytically from the even Fourier transforms $\gamma_{ij}(\omega)$ given in Eqn.~(\ref{Eparftcorrfunc}).
Similarly to the discussion in appendix~\ref{Aliouville_ser} we may obtain the BMS limit analytically by
letting $\tau\to\infty$.

\section{Evaluation of Lamb-Shift Terms}\label{Alambshift}

Given the Fourier transform of the bath correlation functions $\gamma_{ij}(\omega)$, 
we can obtain the odd Fourier transform for the conventions chosen in~(\ref{Efourier}) {\em via} 
\bea\label{Ecpv}
\sigma_{ij}(\omega) &=& \frac{1}{2\pi} \int\limits_{-\infty}^{+\infty} d\bar\omega 
\left[ \int\limits_{-\infty}^{+\infty} d\tau
e^{-i(\omega - \bar \omega) \tau} {\rm sgn}(\tau)\right] \gamma_{ij}(\bar\omega)\nn
&=& \frac{i}{\pi} {\cal P} \int\limits_{-\infty}^{+\infty} \frac{\gamma_{ij}(\bar\omega)}{\bar\omega-\omega} d\bar\omega\,,
\eea
where ${\cal P}$ denotes the Cauchy Principal Value.
Thus, the Lamb shift is equivalent to the exchange field in Refs.~\cite{BRA04,BRA06b}.
With the relation $\tanh(\pi y) = 2/\pi \Im \Psi(1/2 + i y)$ (where $\Psi(z)$ denotes the Digamma function) we can write
\bea\label{Edigammatrafo}
f(\bar\omega)&=&\frac{1}{e^{\beta(\bar \omega-\mu)}+1} = \frac{1}{2}\left[1-\tanh\left(\frac{\beta(\bar\omega-\mu)}{2}\right)\right]\nn
&=& \Im \frac{1}{2} \left[ i + \frac{2}{\pi} \Psi\left(\frac{1}{2} - i \frac{\beta(\bar\omega-\mu)}{2\pi}\right)\right]\,,\nn
1-f(-\bar\omega) &=& 
\Im \frac{1}{2} \left[ i + \frac{2}{\pi} \Psi\left(\frac{1}{2} - i \frac{\beta(\bar\omega+\mu)}{2\pi}\right)\right]\,,
\eea
which is valid for all $\bar \omega$ on the real axis.

Since the integral in Eqn.~(\ref{Ecpv}) goes along the real axis only, we can exploit the above identities in 
the Kramers-Kronig relation
\bea
{\cal P} \int\limits_{-\infty}^{+\infty} \frac{\chi(\omega')}{\omega'-\omega} d\omega' &=& i \pi \chi(\omega)\\
&&+ 2\pi i \sum_{k: \Im \omega_k > 0} \left.{\rm Res} \frac{\chi(\omega')}{\omega'-\omega}\right|_{\omega'=\omega_k}\,,\nonumber
\eea
which is valid whenever the (holomorphic) function $\chi(\omega)$ decays sufficiently fast in the upper complex half plane and
where $\omega_k$ denote the poles of $\chi(\omega)$ in the upper complex half plane.
Now, the imaginary part of the above Kramers-Kronig relation reads
\bea\label{Eintegral}
\Im {\cal P} \int\limits_{-\infty}^{+\infty} \frac{\chi(\omega')}{\omega'-\omega} d\omega' &=&
\pi \Re \chi(\omega)\\
&& + 2\pi \Re \sum_{k:\Im \omega_k > 0} \left.{\rm Res} 
\frac{\chi(\omega')}{\omega'-\omega}\right|_{\omega'=\omega_k}\,,\nonumber
\eea
and can be directly related to Eqn.~(\ref{Ecpv}) when we choose the function
\mbox{$\chi(\omega')=\frac{1}{2} \left[ i + \frac{2}{\pi} \Psi\left(\frac{1}{2} - i \frac{\beta(\omega'-\mu)}{2\pi}\right)\right]\tilde\Gamma(\pm\omega')$}, 
see also Eqn.~(\ref{Edigammatrafo}).
The additional factor $\tilde\Gamma(\pm\omega')$ arises from the (in reality) 
frequency-dependent tunneling rates, compare the Fourier transforms in 
equations~(\ref{Eserftcorrfunc}) and~(\ref{Eparftcorrfunc}), respectively.
For simplicity, we take \mbox{$\Gamma_\alpha(\omega)=\Gamma_\alpha \tilde \Gamma(\omega)$}
and assume a Lorentzian dependence
\bea\label{Elorentz_cutoff}
\tilde\Gamma(\omega) = \frac{\delta^2}{(\omega-\epsilon)^2+\delta^2}
\stackrel{\delta\to\infty}{\Longrightarrow} 1\,.
\eea

\begin{figure}[t]
%\psfrag{PSFomeg}{$\omega$}
%\psfrag{PSFepspid}{$\epsilon+i\delta$}
%\psfrag{PSFepsmid}{$\epsilon-i\delta$}
\includegraphics[width=0.3\textwidth]{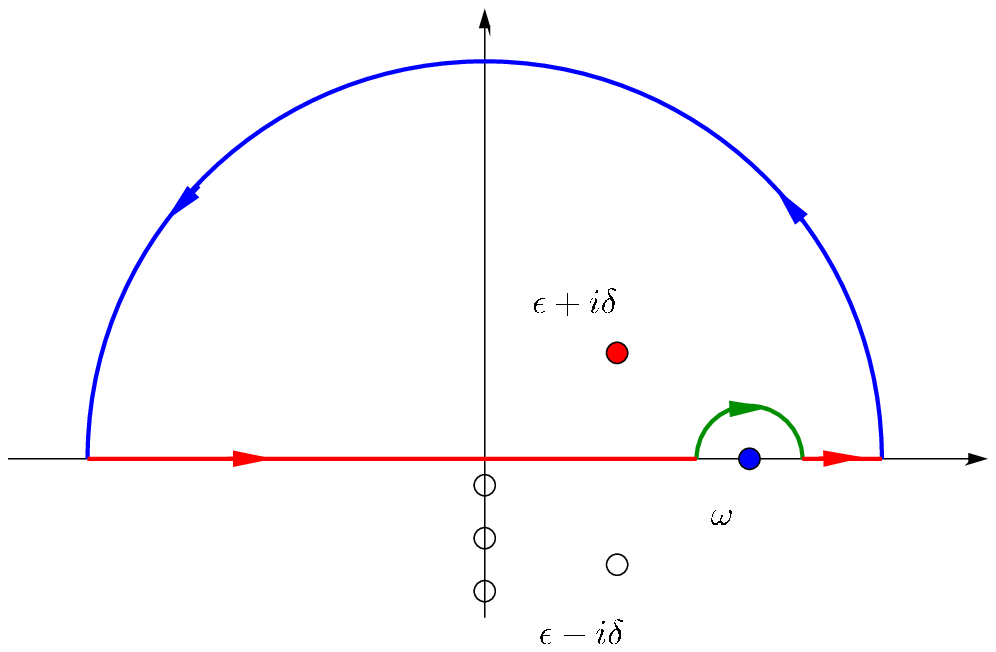}
\caption{\label{Fcontours}[Color Online]
Poles of the function 
\mbox{$\chi(\omega')/(\omega'-\omega)$} with 
\mbox{$\chi(\omega')=\frac{1}{2} \left[ i + \frac{2}{\pi} \Psi\left(\frac{1}{2} - i \frac{\beta(\omega'-\mu)}{2\pi}\right)\right]\tilde\Gamma(\omega')$} 
in the complex plane.
The Digamma function contributes poles on the lower imaginary axis, whereas the Lorentz 
function~(\ref{Elorentz_cutoff}) has two complex conjugate poles of first order, one
of which lies within the integration contour. The integral along the large half circle vanishes due to the Lorentzian cutoff, 
the integral along the real axis
corresponds to the Cauchy Principal value of the left hand side of Eqn.~(\ref{Eintegral}), and the integral along the small 
half circle yields the first term on the right hand side of Eqn.~(\ref{Eintegral}).
The situation can be treated analogously for 
$\Im \tilde\chi(\omega)=[1-f(-\omega)] \tilde\Gamma(-\omega)$.}
\end{figure}

The Digamma function $\Psi(z)$ has poles for non-positive integers. 
Therefore, when $\Im \bar\omega \ge 0$ (the upper complex half plane) the function 
\mbox{$\frac{1}{2} \left[ i + \frac{2}{\pi} \Psi\left(\frac{1}{2} - i \frac{\beta(\bar\omega+\mu)}{2\pi}\right)\right]$}
defined in Eqn.~(\ref{Edigammatrafo}) would behave analytically, such that one would not have to evaluate residues. 
The Digamma function alone would not decay asymptotically, but convergence is ensured by the Lorentzian cutoff
(\ref{Elorentz_cutoff}). 
This Lorentzian cutoff leads to a single pole at $\omega_1=\epsilon + i \delta$ in the upper complex half plane.
In principle, more complicated spectral densities can be fitted by a sum of many Lorentzians \cite{WEL08}.
The situation is depicted in Fig.~\ref{Fcontours}.

With these considerations we can express the Lamb-shift terms in Digamma functions, where the Fourier transforms in 
equations~(\ref{Eserftcorrfunc}) and~(\ref{Eparftcorrfunc}) lead to terms like
\bea
\sigma_a &\equiv& {\cal P} \int\limits_{-\infty}^{+\infty} \frac{f(\bar\omega)\tilde\Gamma(\bar\omega)}{\bar\omega-\omega} d\bar\omega\nn
&=& \Im {\cal P} \int\limits_{-\infty}^{+\infty} 
\frac{\frac{1}{2} \left[ i + \frac{2}{\pi} \Psi\left(\frac{1}{2} - i \frac{\beta(\bar\omega+\mu)}{2\pi}\right)\right]\tilde\Gamma(\bar\omega)}{\bar\omega-\omega} d\bar\omega\nn
&=&
\Re\left\{\Psi\left(\frac{1}{2} - i \frac{\beta(\omega-\mu)}{2\pi}\right)\right\}\tilde\Gamma(\omega)\nn
&&+\frac{1}{2} \Im \left\{
\frac{\delta\left[\pi - 2 i \Psi\left(\frac{1}{2} + \frac{\beta \delta}{2\pi} -i \frac{\beta(\epsilon-\mu)}{2\pi}\right)\right]}
{\delta - i (\epsilon-w)}\right\}\nn
&\stackrel{\delta\to\infty}{\Longrightarrow}&
\Re\left\{\Psi\left(\frac{1}{2} - i \frac{\beta(\omega-\mu)}{2\pi}\right)\right\} - \ln \left(\frac{\beta \delta}{2\pi}\right)
\eea
and also terms like
\bea
\sigma_b &\equiv& {\cal P} \int\limits_{-\infty}^{+\infty} \frac{\left[1-f(-\bar\omega)\right]\tilde\Gamma(-\bar\omega)}{\bar\omega-\omega} d\bar\omega\nn
&=& \Im {\cal P} \int\limits_{-\infty}^{+\infty} 
\frac{\frac{1}{2} \left[ i + \frac{2}{\pi} \Psi\left(\frac{1}{2} - i \frac{\beta(\bar\omega+\mu)}{2\pi}\right)\right]\tilde\Gamma(-\bar\omega)}{\bar\omega-\omega} d\bar\omega\nn
&=&
\Re \left\{ \Psi\left(\frac{1}{2} - i \frac{\beta(\omega+\mu)}{2\pi}\right) \right\}\tilde\Gamma(-\omega)\nn
&&+\frac{1}{2} \Im \left\{
\frac{\delta\left[\pi - 2 i \Psi\left(\frac{1}{2} + \frac{\beta \delta}{2\pi} +i \frac{\beta(\epsilon-\mu)}{2\pi}\right)\right]}
{\delta + i (\epsilon+w)}\right\}\nn
&\stackrel{\delta\to\infty}{\Longrightarrow}&
\Re\left\{\Psi\left(\frac{1}{2} - i \frac{\beta(\omega+\mu)}{2\pi}\right)\right\}- \ln \left(\frac{\beta \delta}{2\pi}\right)\,.
\eea
For nearly flat tunneling rates we observe a logarithmic divergence as $\delta \to \infty$.
These logarithmic counterterms can not always be neglected:
For the parallel configuration, we do indeed observe their cancellation in the Liouvillian, 
whereas for the serial configuration they do not cancel.
In the latter case, the neglect of these terms would even lead to negative density matrices.
We suspect that the reason for this lies in the discrepancy between the localized and the energy-eigenbasis.

\end{appendix}

%%%%%%%%%%%%%%%%%%%%%%%%%%%%%%%%%%%%%%%%%%%%%%%%%%%%%%%%%%%%%%%%%%%%%%%%%%%%%%%%%%%%
%%%%%%%%%%%%%%%%%%%%%%%%%%%%%%%%%%%%%%%%%%%%%%%%%%%%%%%%%%%%%%%%%%%%%%%%%%%%%%%%%%%%
%%%%%%%%%%%%%%%%%%%%%%%%%%%%%%%%%%%%%%%%%%%%%%%%%%%%%%%%%%%%%%%%%%%%%%%%%%%%%%%%%%%%
%%%%%%%%%%%%%%%%%%%%%%%%%%%%%%%%%%%%%%%%%%%%%%%%%%%%%%%%%%%%%%%%%%%%%%%%%%%%%%%%%%%%

\end{document}